# Network Topologies and Dynamics Leading to Endotoxin Tolerance and Priming in Innate Immune Cells

**Yan Fu[1,2], Trevor Glaros[1], Meng Zhu[3], Ping Wang[1], Zhanghan Wu[1,2], John J Tyson[1], Liwu Li[1,*], Jianhua Xing[1,*]**

[1]Department of Biological Sciences and [2]Interdisciplinary PhD Program of Genetics, Bioinformatics and Computational Biology, Virginia Polytechnic Institute and State University, Blacksburg, VA 24060, USA. [3]School of Computing, Clemson University, Clemson, SC 29634, USA.

* Send correspondence to jxing@vt.edu and lwli@vt.edu

Running title: Networks for Endotoxin Tolerance and Priming

# Abstract

The innate immune system, acting as the first line of host defense, senses and adapts to foreign challenges through complex intracellular and intercellular signaling networks. Endotoxin tolerance and priming elicited by macrophages are classic examples of the complex adaptation of innate immune cells. Upon repetitive exposures to different doses of bacterial endotoxin (lipopolysaccharide) or other stimulants, macrophages show either suppressed or augmented inflammatory responses compared to a single exposure to the stimulant. Endotoxin tolerance and priming are critically involved in both immune homeostasis and the pathogenesis of diverse inflammatory diseases. However, the underlying molecular mechanisms are not well understood. By means of a computational search through the parameter space of a coarse-grained three-node network with a two-stage Metropolis sampling approach, we enumerated all the network topologies that can generate priming or tolerance. We discovered three major mechanisms for priming (pathway synergy, suppressor deactivation, activator induction) and one for tolerance (inhibitor persistence). These results not only explain existing experimental observations, but also reveal intriguing test scenarios for future experimental studies to clarify mechanisms of endotoxin priming and tolerance.

## Author Summary

Inflammation is a fundamental response of animals to pathogen invasion. Among the first responders are macrophage cells, which identify and respond to multiple challenges. Their responses must be carefully regulated to kill invading pathogens without causing too much damage to host cells. Excessive activity of macrophages is associated with serious diseases like sclerosis and cancer. Macrophage responses are governed by a complex signaling network that receives cues, integrates information, implements appropriate responses and communicates with neighboring cells. This network must maintain a short-term memory of pathogen exposure. Endotoxin priming is an example. If macrophages are exposed to a small dose of bacterial toxins, they are primed to respond strongly to a second exposure to a large dose of toxin. Endotoxin tolerance, on the other hand, refers to the fact that macrophages are resistant to endotoxin challenges after a large dose pretreatment. The precise molecular mechanisms of both priming and tolerance are still poorly understood. Through computational systems biology, we have identified basic regulatory motifs for priming and for tolerance. Using information from databases and the literature, we have identified molecules that may contribute to priming and tolerance effects. Our methods are generally applicable to other types of cellular responses.

# Introduction

Innate immune cells such as macrophages and dendritic cells constitute the first layer of host defense. Like policemen constantly patrolling the streets for criminal activity, these cells are responsible for initiating the first attack against invading pathogens [1,2]. For example, using Toll-like receptor 4 (TLR4), macrophages recognize lipopolysaccharide (LPS, also called endotoxin), a pathogen-associated molecular pattern (PAMP) that is expressed on the outer membrane of gram-negative bacteria. Within hours of stimulation, hundreds of regulatory genes, kinases, cytokines, and chemokines are activated in sequential waves, leading to a profound inflammatory and anti-microbial response in macrophages [3]. Although effective levels of inflammation require potent cytokine production, excessive or prolonged expression can be detrimental, resulting in various immune diseases, such as autoimmunity, atherosclerosis, sepsis shock and cancers [3,4]. Owing to this double-edged nature of innate immunity, living organisms have evolved a highly complex signaling network to fine-tune the expression of cytokines [5]. A fundamental question in this field is what kinds of network topologies and dynamics in the signaling network ensure the appropriate expression of cytokines. This question is part of a larger current theme in systems biology of the design principles of biological networks. Are there small network motifs that serve as building blocks to perform complex "information processing" functions in biological signaling networks [6-12]? In this context, a systems and computational biology approach may greatly deepen our understanding in innate immunity [13-17].

Here we focus on the signaling motifs responsible for endotoxin priming and tolerance of macrophages. The interaction between host macrophages and bacterial endotoxin is arguably one of the most ancient and highly conserved phenomena in multi-cellular eukaryotic organisms [5]. Through TLR4, LPS activates MyD88-dependent and MyD88-independent pathways, which eventually lead to the regulation of a number of downstream genes and pathways, including the mitogen-activated protein kinase (MAPK), phosphoinositide 3-kinase (PI3K), and nuclear factor κB (NFκB). The integration of these intracellular pathways leads to measured induction of pro-inflammatory mediators. Intriguingly, the induction of inflammatory mediators is also finely controlled by the quantities and prior history of LPS challenges. The latter is physiologically relevant since cells are likely repetitively exposed to stimulants in their natural environment. For example, numerous *in vitro* studies have found that significant induction of cytokine TNF-α and IL-6 requires at least 10 ng/mL LPS in mouse peritoneal macrophages [18,19] and macrophage cell lines [20], and a high dose of LPS (100 ng/mL) is sufficient to trigger a catastrophic "cytokine storm". Strikingly, however, the dose-response relationship can be reprogrammed by two successive treatments with LPS, to give either a reduced or an augmented expression of cytokines (Figure 1A). *In vitro*, preconditioning macrophages with a high dose (HD) of LPS (10−100 ng/mL) renders the cells much less responsive to a subsequent HD stimulation in terms of pro-inflammatory cytokine expression. This phenomenon, known as "endotoxin tolerance" or "LPS tolerance" [21], is reported to last up to 3 weeks *in vivo* [22]. On the other hand, macrophages primed by a low dose (LD) of LPS (0.05−1 ng/mL) show an augmented production of cytokine in response to a subsequent HD challenge, a phenomenon known as “LPS priming” [18,19,23-25]. Both priming and tolerance are present in other cells of the innate immune system including monocytes and fibroblasts, and are highly conserved from mice to humans. Our own studies on murine macrophages show both effects (Figure 1B).

Endotoxin priming and tolerance may confer significant survival advantages to higher eukaryotes. Priming of innate immune cells may enable robust and expedient defense against invading pathogens, a mechanism crudely analogous to vaccination of the adaptive immune system. On the other hand, tolerance may promote proper homeostasis following robust innate immune responses. However, despite these survival advantages, endotoxin priming and tolerance are also closely associated with the pathogenesis of both chronic and acute human diseases. For example, despite the potential ability to limit pro-inflammatory cytokine production, endotoxin tolerance is responsible for the induction of immunosuppression in patients with sepsis shock, and this suppression leads to increased incidence to secondary infections and mortality [22]. Endotoxin priming, on the other hand, reprograms macrophages to super-induction of proinflammatory cytokines. Increasing evidence relates this phenomenon to low-grade metabolic endotoxemia, where an elevated but physiological level of LPS in the host’s bloodstream results in a higher incidence of insulin resistance, diabetes and atherosclerosis [26-29]. Augmented IL-6 expression has also been observed in human blood cells that were primed by LD and challenged by HD LPS [30].

Despite the significance and intense research efforts, molecular mechanisms responsible for endotoxin priming and tolerance are not well understood, apparently due to the complex nature of intracellular signaling networks. Tolerance has been attributed to the negative regulators at multiple levels of the TLR4 signaling pathway. These include signaling molecules (*e.g.* SHIP, ST2, induction of IRAK-M and suppression of IRAK-1), transcriptional modulators (*e.g.* ATF3, p50/p50 homodimers), soluble factors (*e.g.* IL-10 and TGFβ), and gene-specific chromatin modifications [21,31-38]. These negative regulators are likely to work together to drive macrophages into a transient refractory state for cytokine expression after LPS pretreatment [33]. Molecular mechanisms for priming are rarely studied and even less well understood than tolerance. Early studies suggest that like endotoxin tolerance, both intra- and inter-cellular events may be involved in LPS priming [24]. Morrison and coworkers first revealed that LPS priming of cytokine TNF-α production is induced, at least in part, by a reprogrammed counterbalance between endogenous IL-10 and IL-12 in an autocrine fashion [19]. However, it is still elusive exactly how the change in two counteracting soluble secretory products can contribute to the priming effect, and whether LPS priming is exclusively an intercellular event or it takes place at both intra- and inter-cellular levels.

These published observations and our own new experimental results have inspired us to look for all possible mechanisms for LPS priming and tolerance. To do this, we computationally searched the high-dimensional parameter space associated with a generic mathematical model of a three-node regulatory network. The search reveals only three mechanisms accounting for priming (pathway synergy, suppressor deactivation, activator induction) and one for tolerance (inhibitor persistence). Existing experimental results support these mechanisms.

In summary, our approach provides a systematic, quantitative framework for understanding numerous experimental observations, and it suggests new experimental procedures to identify the players and investigate the dynamics of priming and tolerance. Our analysis suggests that endotoxin tolerance and priming are rooted in the basic structure of the immune regulatory network: a signal often triggers synergizing pathways to ensure that sufficient responses can be elicited efficiently, as well as opposing pathways to ensure that the responses can be resolved eventually [2]. Therefore, in addition to shedding light on LPS-induced tolerance and priming, our approach is applicable in the more general context of cross-priming and cross-talk in the signal transduction mechanisms of the innate immune system [39-41].

# Results

### Inducing priming and tolerance in a well-controlled experimental setting

Although separate experimental studies of priming and tolerance have been carried out in many laboratories, no systematic study of both effects has been performed in the same setting. Thus, we first set out to measure priming and tolerance in the same experimental system. We used murine bone marrow derived macrophages (BMDM), which are widely used for measuring LPS responses. BMDM were treated with various combinations of LD (50 pg/mL) and HD (100 ng/mL) LPS for times indicated in Figure 1B. Cells were washed with PBS and fresh medium between consecutive treatments. Figure 1B shows that 50 pg/mL LPS induced negligible *IL-6*, while 100 ng/mL LPS induced robust expression of *IL-6* in BMDM (~3300 fold). Consistent with previous findings, cells pre-treated for 4 h with 50 pg/mL LPS exhibited ~4500 fold induction of *IL-6* when challenged with 100 ng/mL LPS, a ~36% augmentation as compared to cells treated with 100 ng/mL LPS alone ($p < 0.05$). In contrast, cells pretreated for 4 h with 100 ng/mL LPS exhibited only ~700 fold induction of *IL-6* when re-challenged with 100 ng/mL LPS, a ~80% reduction as compared to cells treated with 100 ng/mL LPS alone ($p < 0.05$).

### Identifying motifs that generate priming effect

Figure 1C shows that LPS binding to TLR4 triggers two groups of parallel pathways: MyD88-dependent and (several) MyD88-independent pathways. Together, these pathways control the expression of different but overlapping inflammatory mediators in a delicate time-dependent and dose-dependent manner. Based on these parallel pathways, we proposed a three-node model in Figure 1C as a minimal abstraction of the system. Each node can positively or negatively regulate the activity of itself and the other two nodes. The interactions are governed, we assume, by a standardized set of nonlinear ordinary differential equations (Figure 1C) for $x_j$ = activity of the $j^{\text{th}}$ node ($0 \le x_j \le 1$, $j$ = 1,2,3). For a complete description of the mathematical model, see the section on Materials and Methods. The "network topology" of the model is determined by the sign pattern of the nine interaction coefficients ($-1 \le \omega_{ji} \le 1$, $j,i$ = 1,2,3) which express the magnitude and direction of the effect of node $i$ on node $j$. This is a coarse-grained model, with no distinction between intra- and inter-cellular events. For example, in a real cell the self-regulation of a node may correspond to a feedback loop involving many intermediates, including extracellular cytokines. The simplicity of the model allows full search of the 14-dimensional parameter space (although there are 18 parameters in Table 1, four of them are held constant, as explained in Materials and Methods). Similar three-node models have been studied in other contexts [6,42,43].

We searched the 14-dimensional parameter space of the model for priming and then for tolerance. The behavior of the model is defined as "priming" if the maximum level of the output variable $x_3$ under the priming dose (step 3 in Figure 1A) is small ($x_3 < 0.3$), but with the subsequent high dose (step 4 in Figure 1A) $x_3$ is at least 50% higher than the level reached without priming (step 1 in Figure 1A). Similarly, for "tolerance" the maximum level of $x_3$ must be high enough under the first HD exposure ($x_3 > 0.3$) but less intense by at least 50% under the second HD challenge (step 2 in Figure 1A). Precise criteria for priming and tolerance are provided in Table S1. Brute force search of the parameter space is impractical. Unbiased searching results in <1000 parameter sets exhibiting priming after $10^8$ Monte Carlo steps. Noticing that parameter sets giving priming or tolerance (called "good sets" for convenience) are clustered into a small number of isolated regions in parameter space, we designed a two-stage sampling procedure. First we perform a Metropolis search slightly biased for good sets. Next, to identify any isolated regions of parameter space where good sets are clustered, we analyzed the good sets using K-means clustering and Principal Component Analysis (see Text S1). The good sets then serve as seeds in the second stage of sampling, which restricts Metropolis searching to each local region of good sets. This two-stage procedure allows us to search the parameter space thoroughly and to obtain good-set samples that are large enough for statistical analysis. The overall procedure is illustrated schematically in Figure S1 and discussed in Text S1.

**Three basic mechanisms for the priming effect of LPS**

By trial-and-error, we found that the two experimentally measurable quantities, $\Delta x_1$ and $\Delta x_2$ (see Figure 2A), are effective in dividing the "good" parameter sets into three regions (see Figure 2B). Here $\Delta x_1$ = maximum difference between $x_1$ during the LD priming stage and the steady state value of $x_1$ in the absence of any stimulus, and $\Delta x_2$ = difference between the maximum values of $x_2$ during the HD period with and without the priming pretreatment (Figure 2A). Further analysis (discussed below) revealed that the three groups correspond to three distinct priming mechanisms: "Pathway Synergy" (PS), "Activator Induction" (AI), and "Suppressor Deactivation" (SD). All AI and PS parameter sets show considerable increase in $x_2$ (> 0.1) after the priming stage, while SD does not (Figure S2).

To characterize these priming mechanisms, we next examined the parameter sets within each group for shared topological features. The topology of a regulatory motif is defined as the sign pattern (+, − or 0) of the nine interaction coefficients, $\omega_{ji}$, with the proviso that $\omega_{ji}$'s in the interval [−0.1, 0.1] are set = 0. We define a backbone motif as the simplest network topology that is shared by most of the good priming sets in each group and that is able to generate a priming effect on its own. Therefore, a backbone motif represents a core network structure in each group. Figure 3A shows that each group has its unique backbone motif(s), directly revealing different priming mechanisms in each group.

Figure S3 and Text S1 provide detailed statistical methods used to identify the backbone motifs. The two-dimensional parameter histograms in Figure S4 provide further support for the backbone motifs we have identified.

Figure 3B-D shows typical time-courses and state-space trajectories for the three priming mechanisms (see Table S2 for the parameter values used to generate this figure).

***Pathway Synergy* (*PS*):** As shown in the upper left panel of Figure 3A, the backbone motif of PS mechanism contains both pathways through $x_1$ and $x_2$ activating $x_3$. Under a single HD, the faster pathway through $x_1$ prevents activation of $x_2$, either directly or through $x_3$. Consequently there is no synergy between the two pathways after a single HD. With LD pretreatment, however, $x_2$ is partially activated. During the following HD treatment, this partial activation allows $x_2$ to increase significantly, either transiently (Figure 3B left panel, called "monostable") or persistently (Figure 3B right panel, called "bistable"), despite inhibition from $x_1$ and/or $x_3$. Simultaneous activation of both pathways leads to synergy between them and a priming effect for $x_3$.

***Activator Induction* (*AI*):** In the backbone motif (see upper right panel of Figure 3A), the pathway through $x_1$ (with high activation threshold) inhibits $x_3$, whereas the pathway through $x_2$ (with a low activation threshold) activates $x_3$. Consequently, under a single HD, the two pathways work against each other to prevent full activation of $x_3$. A LD pretreatment partially activates $x_2$ without significantly affecting $x_1$. Then, during the following HD treatment, $x_2$ gets a head start on $x_1$ to induce greater activation of $x_3$ than observed under a single HD. The activation of $x_3$ can be either transient (monostable) or persistent (bistable), as illustrated in Figure 3C and Figure S5A.

***Suppressor Deactivation* (*SD*):** In this case there are two backbone motifs slightly different from each other (the lower panel of Figure 3A). Both motifs contain an inhibition pathway ($x_1$ —| $x_3$) with slow dynamics and low sensitivity to LPS, and an activation pathway ($x_2 \rightarrow x_3$) with fast dynamics and high sensitivity to LPS. The basal level of the suppressor $x_1$ is relatively high, which is typical of some suppressors (*e.g.* TOLLIP, TRAILR, PI3K and nuclear receptors) that are constitutively expressed in macrophages to prevent unwanted expression of downstream pro-inflammatory genes under non-stimulated conditions [44,45]. Compared to AI, in this case the LD pretreatment decreases the level of suppressor $x_1$, through direct inhibition of $x_1$ by $x_2$. The basic SD effect is amplified either by $x_2$ self-activation (backbone motif I) or by negative feedback from $x_3$ to $x_1$ (backbone motif II). As before, the activation of $x_3$ can be either transient (monostable) or persistent (bistable), as illustrated in Figure 3B and Figure S5B.

**Combined backbone motifs may enhance the robustness of the priming effect**

Each of these groups contains many different network topologies (187 in PS, 139 in SD, and 82 in AI). Taking SD as an example, Figure 4A shows the sorted density distribution of the 139 unique topologies represented by the SD parameter sets. The top 7 of these topologies (Figure 4B) comprise 31% of all the SD parameter sets. Consistent with other studies [6,43], the most highly represented topologies contain more links than the corresponding backbone motif, indicating that additional links may increase the robustness of a network. While the two backbone motifs rank Top 27 and Top 10 respectively (Figure 4B), their combination ranks Top 4. The Venn diagram in Figure 4C shows that of the 93% of SD parameter sets that contain at least one of the two backbone motifs, 64% contain both. Notice that the two backbone motifs use different helpers to deactivate the suppressor ($x_1$) under LD, the combination of motifs (Top 4) integrates both helpers so that deactivation of the suppressor can be enhanced (Figure 4C). The results of a similar analysis applied to PS and AI mechanisms are given in Figure S7.

Additionally, in the Figure S8 and Text S1, we discuss a parameter compensation effect that further expands the priming region in the parameter space.

**Slow inhibitor relaxation dynamics is essential for the induction of tolerance**

We used the 3-node model to search for endotoxin-tolerance motifs. The tolerance effect requires that pro-inflammatory cytokine expression ($x_3$) is markedly reduced (by at least 1.5 fold) under two sequential HD treatments with LPS, compared to the level induced by a single HD (see Table S1 for details). Over 1660 unique topologies are found to give a tolerance effect (Figure 5A), indicating that the requirements for tolerance are much lower than for priming. A typical time course (Figure 5B, left panel) highlights the essential dynamical requirement for tolerance — to sustain a sufficiently high level of inhibitor ($x_1$ in this case) after the first HD of LPS so that $x_3$ is less responsive to the second HD stimulus. The effect is transient: if the second HD stimulus is delayed long enough for the suppressor to return to its basal level, then the tolerance effect is lost (Figure 5B, right panel). This “memory” effect has been noticed in other modeling studies [46-49] and is consistent with experimental observations. For example, the tolerance status of IL-6 is reported to persist for 48 h after the initial HD of LPS, but beyond this time a re-challenge started to recover the expression of IL-6 [34]. Figure 5C shows two backbone motifs that support temporary persistence of the inhibitor: by slow removal or by positive auto-regulation of the inhibitor.

**The dosing scenarios for priming and tolerance are well separated**

It is of interest to ask whether priming and tolerance can be observed in a single 3-node network given the corresponding dosing conditions. It turns out that about 11% of the priming motifs exhibit tolerance as well, and most of them belong to the SD or the AI mechanism. Figure 6A shows qualitatively the dose-response relationship for priming and tolerance in a typical network motif. First, both priming and tolerance require a relatively large second dose (>0.5). Second, the dosing regions for priming and tolerance are well separated. A low first dose (0.1−0.4) leads to priming while a higher one (0.5−1) leads to tolerance. There exists a range separating the priming and the tolerance region where neither are observed.

**Signaling durations affect the induction of priming and tolerance**

Most experimental studies of priming and tolerance are performed with fixed durations of the three time periods ($T_1$, $T_2$, and $T_3$ in Figure 1A). Time-course measurements are rarely reported. The phase diagrams in Figure 6B & C show how varying each time period can affect the induction of priming and tolerance in a typical network motif.

Altogether, these results reveal important dynamical requirement in priming and tolerance and suggest systematic studies in real biological experiments.

The left panel of Figure 6B shows the effects of varying stimulus durations ($T_1$ and $T_3$) at fixed gap duration ($T_2$). To generate priming, $T_1$ must be sufficiently long, while $T_3$ can be relatively short (left panel of Figure 6B). A sufficient priming duration is crucial because the system utilizes this time to activate/deactivate the regulatory pathway with slower dynamics, *i.e.*, the synergizing pathway in PS and the suppressor pathway in SD. Therefore, if $T_1$ is too short, one may erroneously conclude that priming does not exist in the system. On the other hand, tolerance is less dependent on $T_1$ (right panel of Figure 6B).

Figure 6C shows results when all durations are varied under the constraint $T_1 = T_3$. In this case, both priming and tolerance require that $T_2$ is sufficiently short compared to the time required for the system to relax to its basal state after the first stimulus. This result reveals priming and tolerance as essentially the result of cellular memory of the first stimulation.

# Discussion

Using a simple yet flexible model of cellular signaling pathways, we have carried out a systematic study of the topological and dynamic requirements for endotoxin priming and tolerance in cells of the innate immune system. Our study reveals that the phenomena of priming and tolerance can be attributed to a few characteristic network motifs (called “backbone” motifs) that are simple yet effective combinations of feed-forward loops, negative feedback signals, and auto-activation. In addition to reconciling the limited available experimental data on endotoxin priming and tolerance, our models suggest novel, testable hypotheses regarding the molecular mechanisms responsible for these effects.

### Essential modalities for priming and tolerance

Our *in silico* analysis identifies three basic mechanisms for priming (Figure 7). In these mechanisms two pathways interact either constructively (pathway synergy−PS) or destructively (activator induction−AI, suppressor deactivation−SD). Compared to the response of these systems to a single high dose (HD) of LPS, a priming dose of LPS modifies the relative phases of the two pathways so as to strengthen pathway synergy (for PS mechanism) or weaken pathway interference (for SD and AI mechanisms).

In this work we define the priming effect as a response of $x_3$ that is at least 50% higher with priming than without. The threshold of 50% is consistent with experimental observations [23,25], but to be sure that our conclusions are robust, we also performed the computational analysis at two other thresholds: 30% augmentation or 70% augmentation (i.e., $\lambda$=1.3 or $\lambda$=1.7 in Table S1). In both cases we obtained results similar to those shown in Figure 2B, corresponding to the three priming mechanisms, although the exact percentage of each priming mechanism among the data sets varies with the priming threshold.

The priming effect may be viewed as a primitive counterpart of the more sophisticated memory mechanisms of the adaptive immune system. For a limited period of time after exposure to a weak stimulus, the system is prepared to launch a stronger response to a second exposure to the (same or another) stimulus [39,50]. On the other hand, tolerance reflects a transient refractory status to produce inflammatory cytokines due to the memory of an earlier exposure.

**Supporting experimental evidences at intra- and inter-cellular levels**

The actual molecular and cellular networks responsible for endotoxin priming and tolerance are highly complex, involving both intra- and inter-cellular signaling modalities. A combination of priming/tolerance motifs most likely coexist in real signaling networks, and their interactions will determine the specific properties of the priming/tolerance effect *in vivo*. LPS is known to activate multiple intracellular pathways through TLR4, including MyD88-dependent, TRIF-dependent pathways [51]. Cross-talk among these pathways may be differentially modulated by low *vs.* high dosages of LPS, and thus contribute to differential priming and tolerance [37,52,53].

Endotoxin tolerance has drawn significant attention in the past due to its relevance to septic shock. Existing literature reveals the involvement of multiple negative regulators (SHIP, ST2, IL-10, IRAK-M, SOCS1) at either intracellular or intercellular levels. Many of them are shown to be persistently elevated during endotoxin tolerance, a key feature (confirmed by our systems analysis) creating a refractory state that suppresses the expression of pro-inflammatory mediators (see Table 2). For example, SHIP and ST2 are documented to have very slow degradation rates. On the other hand, negative regulators with faster turn-over rates, such as A20 and MKP1 (induced between 2−4 h by LPS), are known not to be required for LPS tolerance [21,54].

In terms of priming, our *in silico* results are consistent with limited experimental data regarding potential molecular mechanisms. For example (Figure 8A), IL-12 and IL-10 are differentially induced by low *vs.* high dose LPS, and subsequently serve as autocrine mediators to modulate LPS priming [19]. Figure 8B provides a second example. Low dose LPS (50 pg/mL) can selectively activate transcription factor C/EBPδ, yet fails to activate the classic NFκB pathway [53]. Hence, by a pathway synergy motif, the selective activation of C/EBPδ by low dose LPS may synergize with NFκB under the subsequent high dose to induce the priming effect. While the removal of nuclear repressor by low dose LPS is reported [53], further evidence for the predicted suppressor deactivation mechanism awaits additional, targeted experimentation. In this context, one needs to be aware that our predicted network motifs are simple topologies that have the potential to generate priming or tolerance, within proper parameter ranges. Our predictions warrant further experimental studies to determine the physiologically relevant ranges of signaling parameters required for priming and tolerance.

Our analysis of priming and tolerance is not limited to LPS. Bagchi *et al.* showed that cross-priming may happen between specific TLRs [41]. Ivashkiv and coworkers reported that IFN-γ can prime macrophage for an augmented response to a variety of stimulants, including bacterial LPS, virus, IFN-α/β and IFN-γ itself [39,40]. IFN-γ self-priming is similar to LPS self-priming: a low dose can prime for boosted expression of interferon-responsive genes. The priming mechanism as reported by Hu *et al.* resembles the AI strategy [55]. Interferon-responsive genes such as IRF1 and IP-10 are transcriptionally induced by transcription factor STAT1, and are inhibited by SOCS1 through a negative feedback mechanism. Low dose IFN-γ (1 U/ml) is able to elevate the expression level of STAT1, preparing macrophage for a boosted activation of STAT1 (through phosphorylation and dimerization of STAT1) under the high dose IFN-γ stimulation. With STAT1 being active, however, the inhibitor SOCS1 cannot be expressed during the priming stage, resulting in an augmented expression of IRF-1 and IP-10 (Figure 8C). Furthermore, Figure 8C suggests a possible cross-priming between IFN-γ and TLR4 via a PS mechanism. Priming of macrophage by a low dose IFN-γ promotes STAT1 expression, which may synergistically cooperate with NFκB to give boosted cytokine expression to secondary stimulation by LPS [55,56]. Further experimental studies are needed to confirm the prediction.

**Limitations of three-node models and further theoretical studies**

Three-node models have been used to analyze functional network motifs in several contexts [6,7,43]. The simplicity of three-node models allows a thorough search of the parameter space. However, the model should be viewed as a minimal system. A typical biochemical network surely has more than three nodes. Therefore each node or link in the three-node model is normally coarse-grained from more complex networks. The model parameters are also composite quantities. Three-node models are limited in their ability to generate certain dynamic features such as time delays. Figure 3A shows the backbone motifs of the three mechanisms we have identified. Further studies of models with additional nodes will be necessary to determine whether all of the links are necessary. For example, in Figure 8B, we cannot find evidence for IL-6 inhibiting C/EBPδ (either by direct or indirect links). This lack of evidence may indicate a missing link waiting for experimental confirmation, or it may indicate a limitation of the three-node model. The parameter search algorithm developed in this work can be applied to models with 4 or more nodes, although the search space grows rapidly with the number of nodes.

Despite the above-mentioned limitations, we expect that the three priming mechanisms and the one tolerance mechanism discovered here are quite general, holding beyond the three-node model. We expect that the present work can serve as a basis for analyzing larger networks with more mechanistic details. As illustrated in Figure 8, motifs can be combined together in series or in parallel, and these combined structures may lead to new dynamic properties of functional importance.

**Suggested experimental design**

Our analysis in Figure 6 suggests that systematic studies of signal durations ($T_1$, $T_2$ and $T_3$) may reveal important details of the dynamics of priming and tolerance. For example, both relatively short (4 h, as the experiment in this paper) and longer priming duration ( $\geqslant$ 20 h) are exhibit priming effects in macrophages [25]. Relatively fast transcriptional regulators like NFκB and AP-1, as well as numerous signaling repressors such as PI3K and nuclear receptors, may be involved in intracellular priming motifs, inducing priming in response to short pretreatments. On the other hand, a longer pretreatment orchestrates more complex intercellular pathways whereby autocrine or paracrine signaling of cytokines (e.g. IL-10, IL-12 and type I IFNs) might dominate the induction of priming effects [19]. Therefore, measurements of the full time spectrum are necessary to reveal different parts of the network contributing to priming/tolerance.

Furthermore, our analysis predicts that priming networks may respond in two distinct fashions: monostable (transient super-induction of cytokine) or bistable (sustained super-induction of cytokines). Time-course measurements can distinguish between these two responses, keeping in mind that the bistable behavior predicted here is relative to the effective time-scale of the model. Each motif considered here is embedded in a larger network. Eventually, in a healthy organism pro-inflammatory cytokines have to be cleared out by some other slow processes that resolve the inflammation. On this longer time scale, the sustained induction of cytokines predicted by some of our models would be resolved.

The analysis presented in Figure 2B suggests a plausible hypothesis to characterize underlying mechanisms of endotoxin priming. High throughput techniques can be used to identify genes and proteins that are significantly changed by low dose pretreatment. Likely candidates can be assayed during the course of a priming experiment, and the time-course data analyzed as in Figure 2B to identify the critical regulatory factors.

Our analyses and simulations reveal that the priming effect is quite sensitive to system dynamics, *i.e.*, to parameter values and initial conditions. It is well documented that many biological control systems, especially those involving gene expression, are stochastic in nature. Consequently a population of seemingly identical cells may respond heterogeneously to a fixed experimental protocol. In this case, single-cell measurements may reveal cell-to-cell variations in priming and tolerance responses [57-59].

Taken together, our integrated and systems analyses reconcile the intriguing paradigm of priming and tolerance in monocytes and macrophages. Given the significance and prevalence of this paradigm in immune cells to diverse stimulants other than LPS, our identified functional motifs will serve as potential guidance for future experimental works related to macrophage polarization as well as dynamic balance of immune homeostasis and pathogenesis of inflammatory diseases.

# Materials and Methods

### Mathematical description

The following mathematical formalism is used to describe the dynamics of the three-node system,

$$\frac{\mathrm{d}x_j}{\mathrm{d}t} = \gamma_j (G(\sigma_j W_j) - x_j)$$

where $G(a) = \frac{1}{1+e^{-a}}$, and $W_j = \omega_{j0} + \sum_{i=1}^{3} \omega_{ji} x_i + S_j$. Notice that $x_j(t)$ lies between 0 and 1 for all *t*. All variables and parameters are dimensionless. $G(\sigma_j W_j)$ is a generic "sigmoidal" function with steepness (slope at $W_j = 0$) that increases with $\sigma_j$. Each $\omega_{ji}$ is a real number in [-1, 1] with its absolute value denoting the strength of the regulation; $\omega_{ji} > 0$ for the "activators" and $\omega_{ji} < 0$ for "inhibitors" of node *j*. The sum, $W_j$, is the net activation or inhibition on node *j*, and $\omega_{j0}$ determines whether node *j* is "on" or "off" when all input signals are 0. The parameters $\gamma_j$ determine how quickly each variable approaches its goal value, $G(\sigma_j W_j)$ for the present value of $W_j$. Because the magnitudes of the weights are bounded, $|\omega_{ji}| < 1$, it is possible to do a thorough and systematic search of all possible weight matrices, even for networks of moderate complexity, *e.g.*, *K* (= number of non-zero $\omega_{ji}$'s) < 20. The formalism is close to that used by Vohradsky [60,61] and others [62,63] previously. More detailed discussions and applications of the formalism can be found in [64-66].

The model contains 18 parameters: 9 $\omega_{ji}$'s, 3 $\gamma_j$'s, 3 $\sigma_j$'s and 3 $\omega_{j0}$'s. By setting $\gamma_3 = 1$, we fix the time scale of the model to be the response time of the output variable, $x_3(t)$. We set $\omega_{30} = -0.50$, so that the response variable is close to $x_3 = 0$ in the absence of input. We also chose $\sigma_3 = 6$ as a moderate value for the sigmoidicity of the output response. Apart from that, $\omega_{20}$ is set to be $-0.25$ so that the $x_2$ pathway is responsive to LD stimulation.

**Monte Carlo sampling algorithm**

Our goal is to sample points in a 14-dimensional parameter space that is bounded and continuous. The sampling algorithm needs to search the parameter space thoroughly and generate sample parameter sets that are statistically unbiased and significant. Our strategy is a random walk based on the Metropolis Algorithm [67] through parameter space according to the following rules:

0. Choose an initial parameter set $\boldsymbol{\theta}_0$ and determine its score: $\Omega_0 = 0$ if it is a "good" set, or $\Omega_0 = 1$ if it is not. (See Text S1 for the definition of a good set of parameters for priming or for tolerance.)

1. Generate parameter set $\boldsymbol{\theta}_{k+1}$ from $\boldsymbol{\theta}_k$ by $\boldsymbol{\theta}_{k+1} = \boldsymbol{\theta}_k + \lambda \boldsymbol{\zeta}$, where $\lambda = 0.025$ specifies the maximum displacement per step, and $\boldsymbol{\zeta}$ is a vector of random numbers with uniform distribution between -0.5 and 0.5.

2. Compute $\Omega_{k+1}$. If $\Omega_{k+1} \le \Omega_k$, then accept the step from *k* to *k*+1. If $\Omega_{k+1} > \Omega_k$, then accept the step from *k* to *k* + 1 with probability $\rho$. Otherwise, reject the step *k* to *k*+1.

3. Update *k*. If *k* is larger than a maximum step number, stop. Otherwise return to step 1.

We pursue this strategy in two stages. In stage 1, we set $\rho = 0.0025$ (see Text S1), so that the random walk has larger tendency to stay in "good" regions of parameter space, but can also jump out of a good region and searches randomly until it falls into another good region (which may be the same region it left). Stage 1 generates a random walk of $10^9$ steps, which is sampled every 100 steps. From this sample of $10^7$ parameter sets only the good ones are saved, giving a sample of $\sim 8\times10^4$ good parameter sets. These data are then analyzed as described below:

1. The K-means algorithm is applied to identify possible clusters of good parameter sets in the 14-dimensional parameter space. The clustering result is then visualized through the first two principal components (which account for ~60% of the data variance) under Principal Component Analysis.

2. One parameter set is chosen from each possible cluster to serve as starting points for stage 2.

Stage 2 is a repeat of stage 1 with $\rho$ = 0. In this case the random walk never leaves a good region. The purpose of stage 3 is to generate a large sample of good parameter sets that may occupy different regions of parameter space. The random walks are sampled every 100 steps, generating $10^6$ good parameter sets from each starting point. Each parameter set must pass an additional test for "biological relevance" (see Text S1 for details) before further analysis.

While the results reported in the main text are from one run of the search procedure, the whole procedure was repeated several times with random initial starting point in stage 1. The final results of these repeated runs agree with each other, confirming the convergence of our search procedure.

**Discretization of continuous parameter matrix into topology matrix**

In order to analyze the topological feature of each priming/tolerance mechanism, one needs to map the continuous parameters $\omega_{ji}$ into a discretized topological matrix $\tau_{ji}$. In the topological space, variables are only described by (−, 0, +) representing inhibition, no regulation and activation, respectively. A cut off value (= 0.1) is used to perform the discretization, following the rules below:

$$\tau_{ji} = \begin{cases} -1, & if \;\; \omega_{ji} \le -0.1 \\ 0, & if \;\; -0.1 < \omega_{ji} < 0.1 \\ 1, & if \;\; \omega_{ji} \ge 0.1 \end{cases}$$

**Experimental studies of LPS priming and tolerance**

Murine bone marrow derived macrophages from C57BL/6 wild type mice were harvested as described previously [53]. Cells were cultured in DMEM medium (Invitrogen) supplemented with 100 units/mL penicillin, 100 µg/mL streptomycin, 2 mM l-glutamine, and 10% fetal bovine serum (Hyclone) in a humidified incubator with 5% $CO_2$ at 37 ℃. Cells were treated with LPS (*E. coli* 0111:B4, Sigma) as indicated in the figure legend.

RNAs were harvested using Trizol reagent (Invitrogen) as previously described [53]. Quantitative real-time reverse-transcription (RT)-PCR were performed as described [68]. The relative levels of *IL-6* message were calculated using the $\Delta\Delta$Ct method, using *GAPDH* as the internal control. The relative levels of mRNA from the untreated samples were adjusted to 1 and served as the basal control value.

## Acknowledgements

We thank Dr Xiaoyu Hu for helpful discussions.

## Supporting Information

Supporting Information includes the following files:

1. Supporting text (1 text file):

- **Text S1**. Detailed explanation of parameter search algorithm, modeling methods and statistical analysis of motifs.

2. Supporting figures (8 figures):

- **Figure S1**. Illustration of the two-stage Metropolis search procedure.
- **Figure S2**. Distribution of changes in the initial condition of x2 between primed and non-primed system.
- **Figure S3**. Statistical method used to identify backbone motifs from priming and tolerance data.
- **Figure S4**. Parameter correlations highlight the backbone motifs of each priming mechanism.
- **Figure S5**. Typical time course and corresponding trajectory in the phase space.
- **Figure S6**. Change in the robustness rank as a result of variations in the topology cut-off.
- **Figure S7**. Topologies of the PS and AI mechanisms.
- **Figure S8**. Parameter correlation and compensation affects the robustness of the model.

3. Supporting tables (3 tables):

- **Table S1**. Criteria identifying priming and tolerance for a given parameter set.
- **Table S2**. Parameter sets used to generate time course and phase-space trajectory in Figure 3 and Figure S5.
- **Table S3**. Experimental literatures supporting the network details in Figure 8.

# Tables

**Table 1.** Description of modeling parameters.

| **Parameter** | **Description** |
|---|---|
| $x_j$ | Concentration (or activity) of species *j* |
| $\gamma_j$ | Time scale of $x_j$ dynamics |
| $\omega_{ji}$ | Regulation strength of $x_i$ on $x_j$ |
| $\omega_{j0}$ | Activation threshold of $x_j$ |
| $\sigma_j$ | Nonlinearity of the regulation relation associated to species $x_j$ |
| $S_j$ | External signal strength acting on $x_{j.}$ ($S_3$=0, $S_1$=$S_2$) |

**Table 2**. Experimental evidence supporting the proposed tolerance mechanism.

| **Molecular Candidate** | **Inhibition Target** | **Persistent Strategy** | **Reported Evidence** | **Reference** |
|---|---|---|---|---|
| IRAK-M | IRAK-1 and IRAK-4 signaling | Slow time scale | Both mRNA and protein level of IRAK-M kept increased until 24 h with LPS stimulation. | [31] |
| SHIP | NFκB pathway | Slow time scale; Positive auto-regulation of upstream regulator | Slow but sustained production of SHIP (peaked at 24 h and remained high until 48 h with LPS stimulation), regulated via autocrine-acting TGF-β; long half-life of SHIP protein. | [33] |
| SOCS1<br><br>(under debate) | IRAK and NFκB pathway | Slow time scale | SOCS1 mRNA levels remains detectable 24 h post LPS stimulation. | [69] |
| ST2 | MyD88 and Mal | Slow time scale | ST2 is induced at 4 h and lasts until 48 h with LPS stimulation. | [32] |
| IL-10<br><br>(required but not necessary for tolerance) | MyD88-dependent pathway (IRAK, TRAF6) | Slow time scale;<br><br>Positive autoregulation | Significant level of IL-10 was detected with prolonged (24 h) LPS stimulation, and the level is sustained until 48 h. The IL-10-activated STAT3 is required for efficient induction of IL-10. | [35,70-72] |
| DNA methylation and chromatin remodeling | Proinflammatory cytokine (TNF-α) gene expression | Slow time scale | Sustained methylation of H3 (lys9), increased and sustained binding of RelB (as transcriptional repressor) on TNF-α promoter in tolerant THP-1 cells. | [36,73] |

## Figures

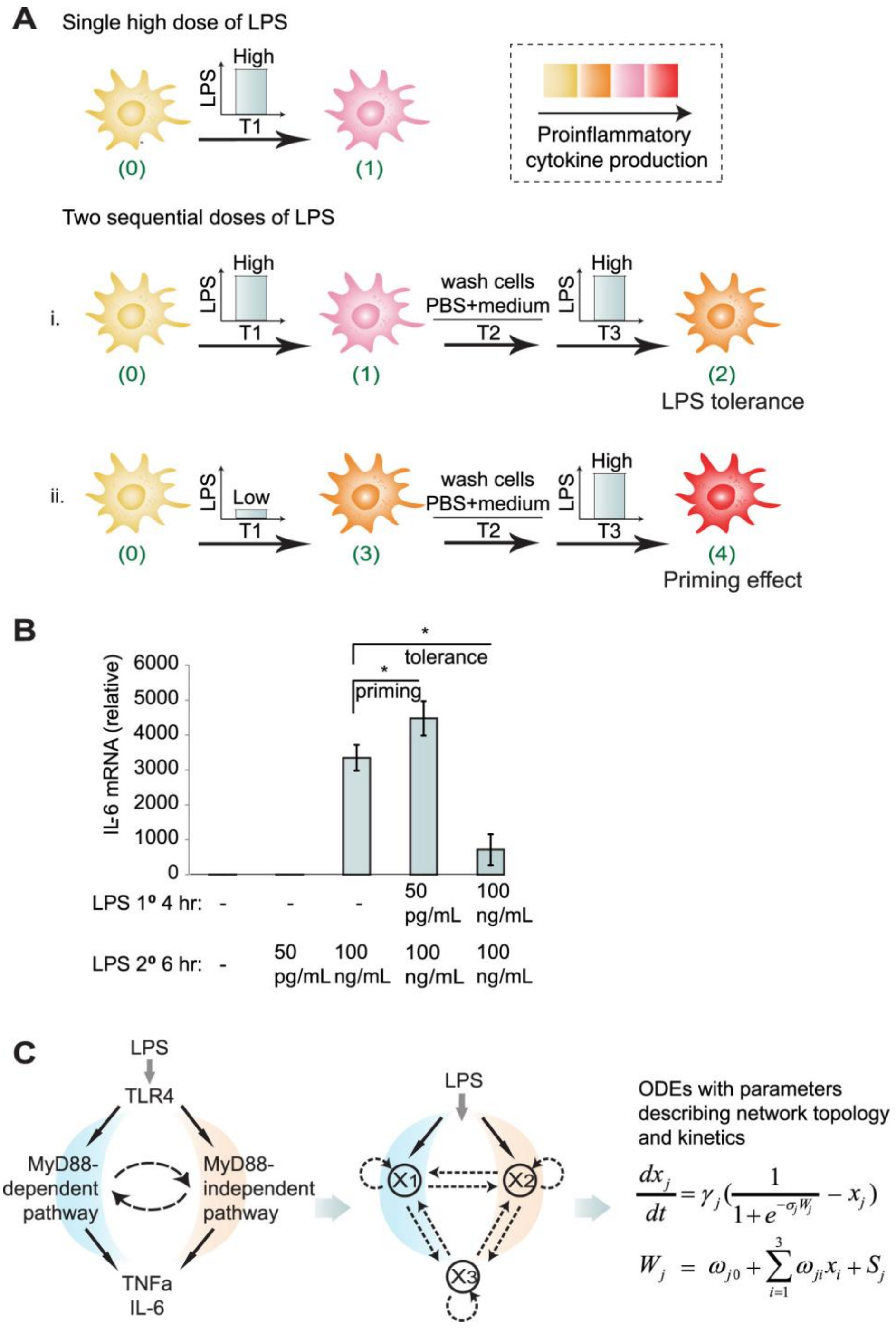

$$\frac{dx_j}{dt} = \gamma_j \left( \frac{1}{1+e^{-\sigma_j W_j}} - x_j \right)$$

$$W_j = \omega_{j0} + \sum_{i=1}^{3} \omega_{ji} x_i + S_j$$

**Figure 1.** Formulation of the problem. (A) Schematic illustration of *in vitro* experimental studies of LPS-induced tolerance and priming effect in macrophages. (B) *IL-6* mRNA

levels of murine bone marrow derived macrophages treated with various combinations of LPS. * $p<0.05$. (C) Abstraction of the parallel LPS associated pathways into a three-node network motif and the corresponding mathematical model based on ordinary differential equations. Refer to Materials and Methods for details.

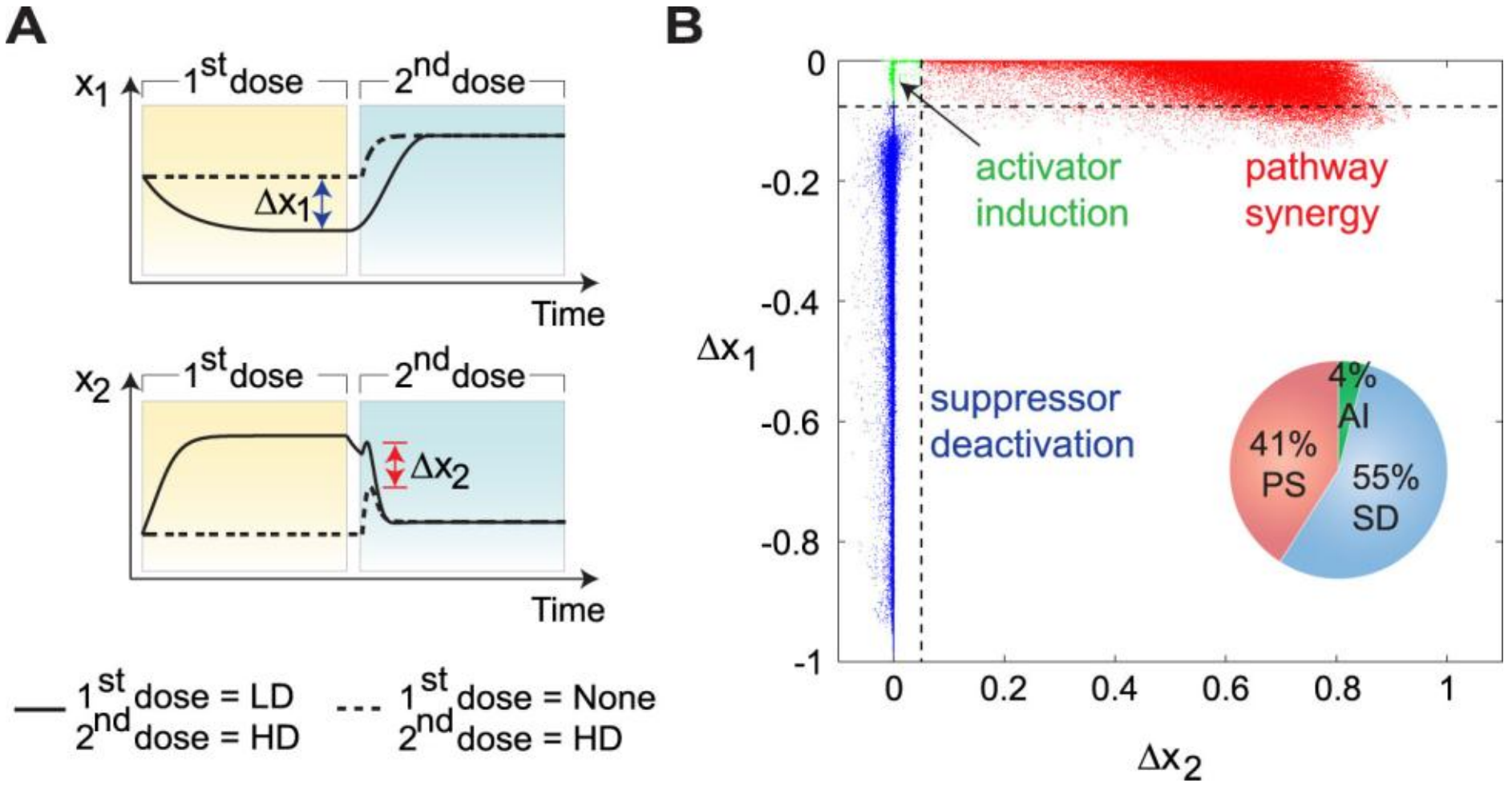

**Figure 2.** Three priming mechanisms revealed by time-course patterns. (A) Definition of clustering axis $\Delta x_1$ and $\Delta x_2$. $\Delta x_1$ refers to the maximum difference between $x_1$ during the LD priming stage and the steady state value of $x_1$ in the absence of any stimulus. $\Delta x_2$ refers to the difference between the maximum values of $x_2$ during the HD period with and without priming pretreatment. (B) The time courses of the priming data sets naturally divide into three clusters, corresponding to three priming mechanisms. The pie chart shows the relative frequencies of the priming mechanisms among all the priming parameter sets.

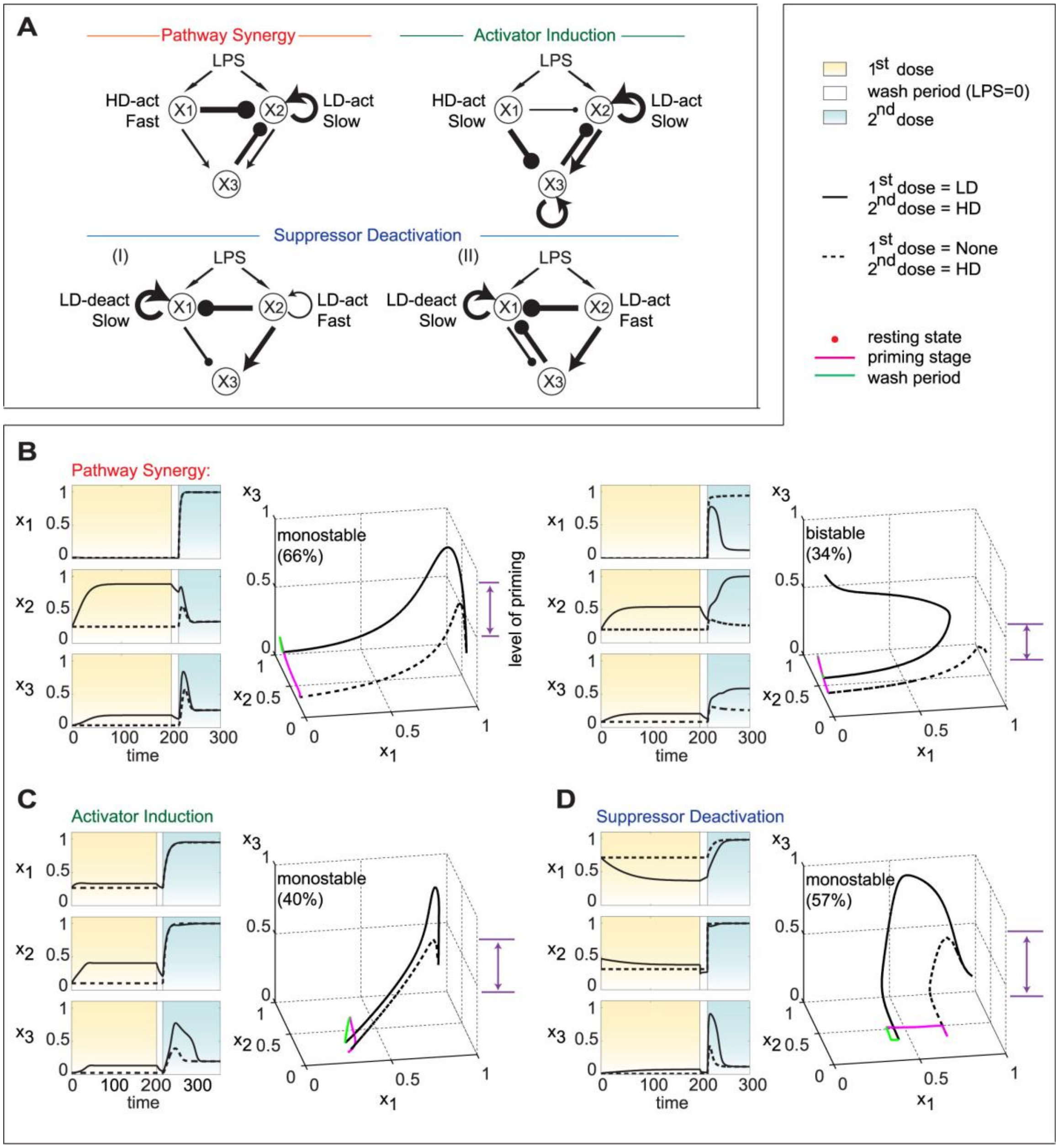

**Figure 3.** Details of the three priming mechanisms. (A) Backbone motifs (topological features shared by most of the good parameter sets) of each priming mechanism (see Figure S3 and Text S1 for details). The width of a line is proportional to the mean value of the corresponding $\omega_{ji}$ among data sets under each priming mechanism. The “slow” and “fast” time scales reflect the values of $\gamma_j$ in comparison to $\gamma_3 = 1$. (B-D) Typical time courses and corresponding phase space trajectories with or without LD pretreatment. Bistable results for AI and SD are shown in Figure S5.

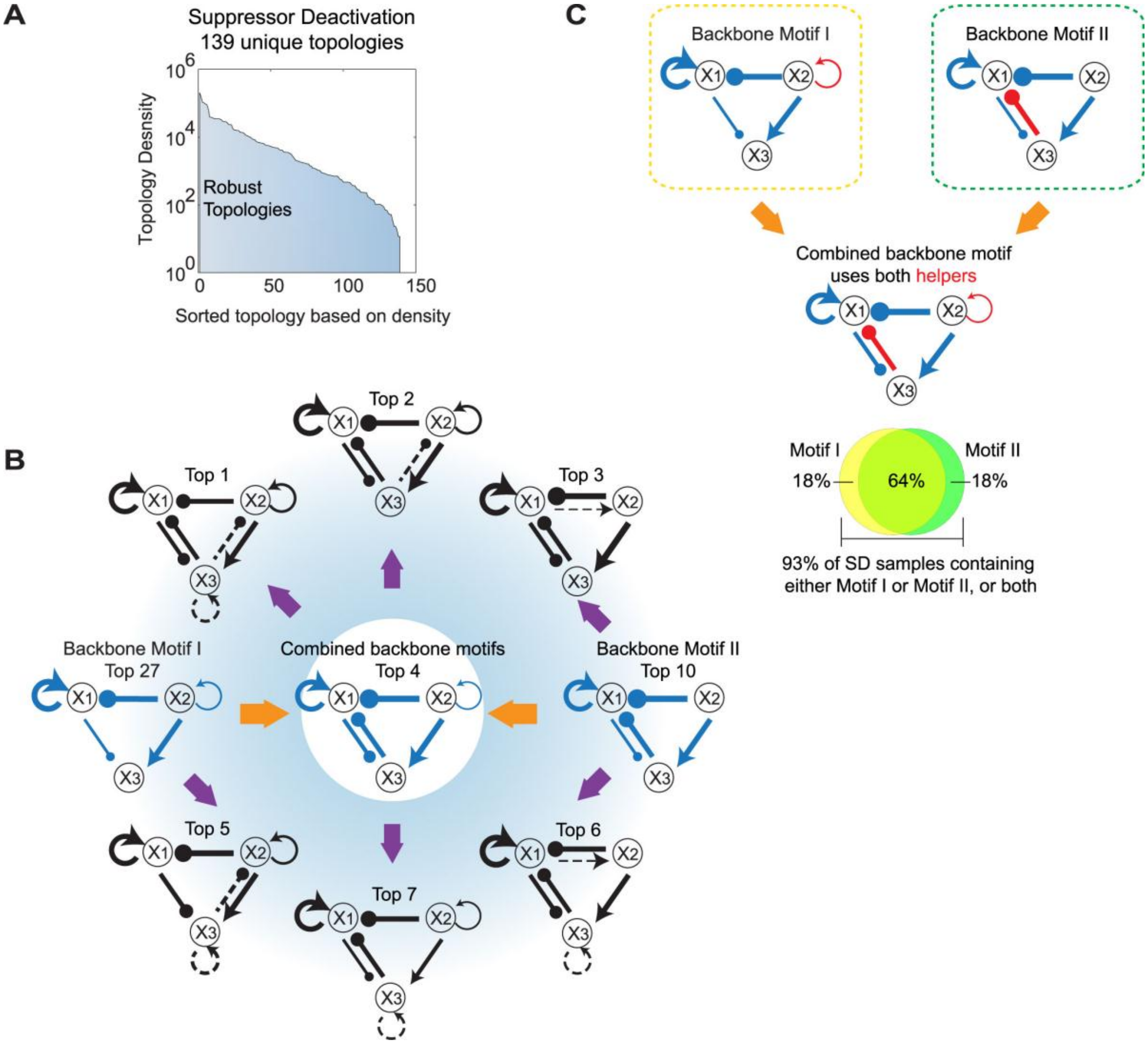

**Figure 4.** Analysis of the robust priming topologies in the SD mechanism. (A) 139 unique topologies under SD mechanism sorted by topology density (see Figure S6 and Text S1 for detailed discussion). (B) The highest seven density topologies and the backbone motifs. Line widths are proportional to the mean value of samples of the corresponding topology. Dashed lines denote the additional link present in the top topologies but absent in the backbone motif. (C) Combination of the two backbone motifs is common in the SD data sets. 93% of SD data sets are found to contain either Motif I or Motif II as the backbone motif. Among them, 64% contain both Motif I and Motif II.

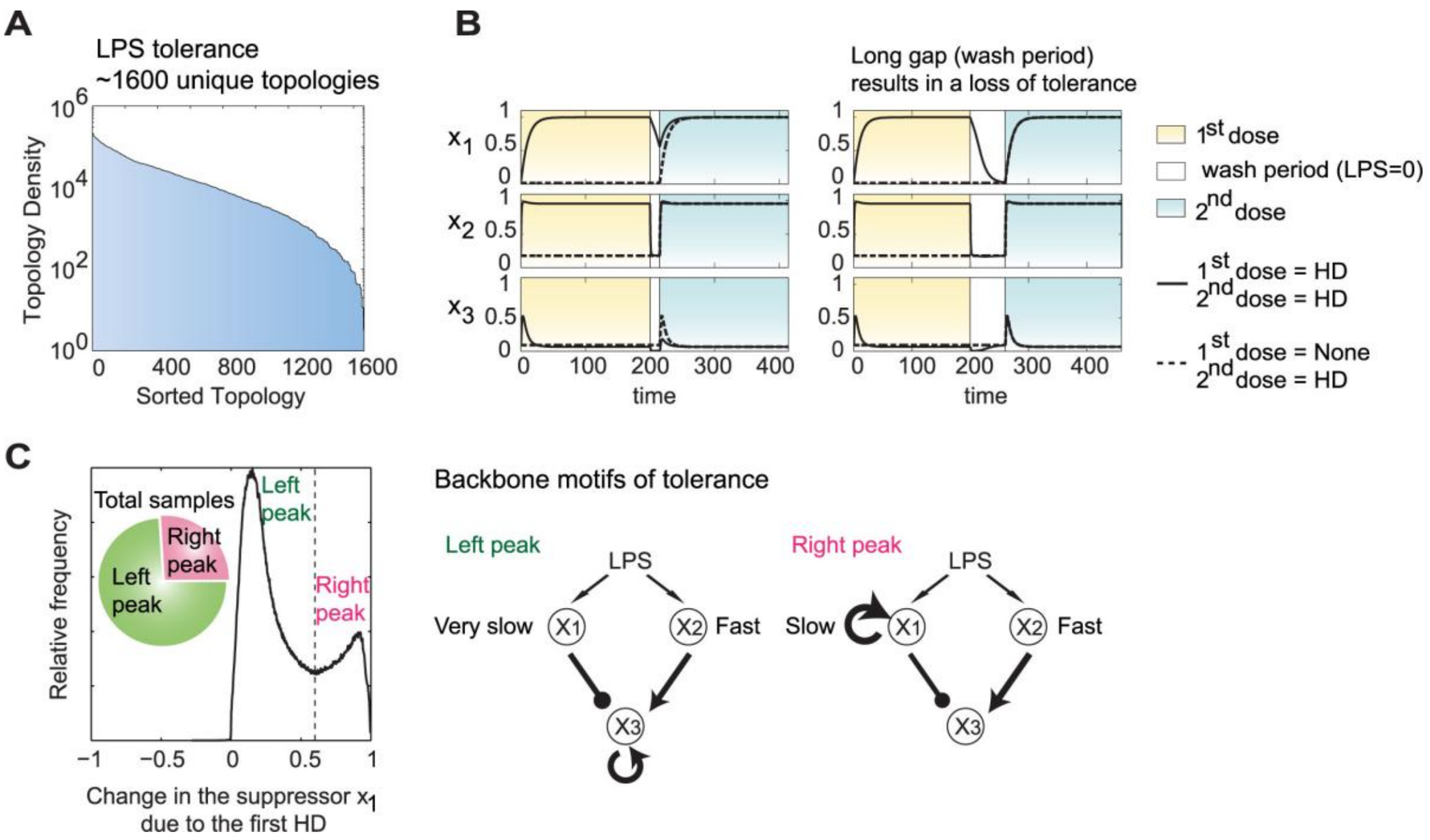

**Figure 5.** Analysis of the tolerance data sets. (A) The unique topologies generating a tolerance effect sorted by topology density. (B) Typical time courses shown with normal (left panel) or elongated (right panel) gap period between the two doses. Solid line: time course tracking the dynamics of the system under the first HD stimulation, in gap period and under a second HD stimulation. Dashed line: time course tracking the dynamics under a single HD treatment; in this case the system is treated with no LPS during the otherwise first HD period. (C) Distribution of the change of $x_1$ level due to the initial HD stimulation reveals two mechanisms to achieve slow relaxation dynamics in the inhibitor (left panel) and the corresponding two backbone motif (right panel).

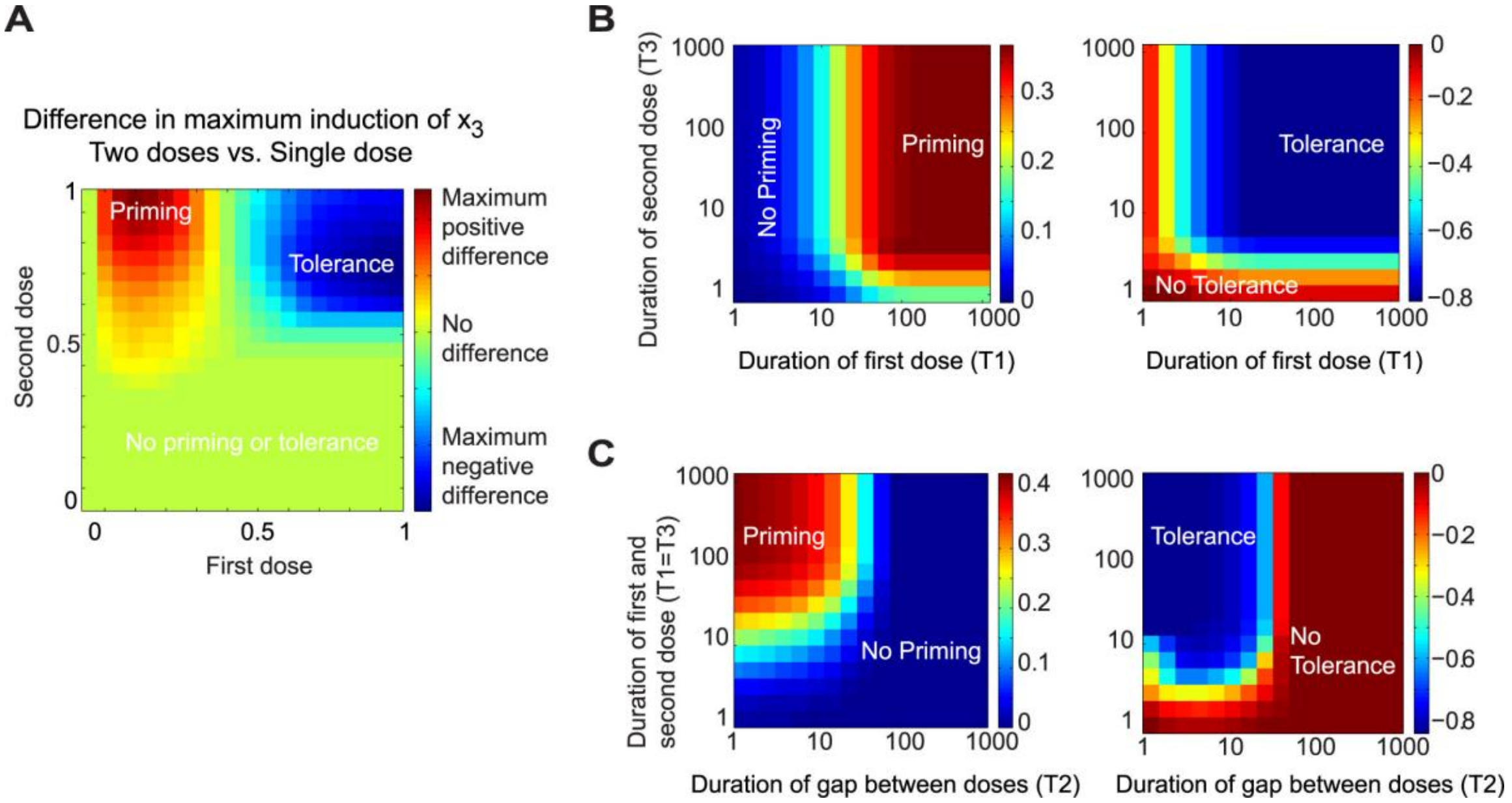

**Figure 6.** Phase diagrams for priming and tolerance in a typical network motif. (A) Regions of dosing conditions for tolerance and priming are well separated. (B) Both priming and tolerance effects are affected by the duration of two sequential treatments (with the gap period between two doses being fixed). (C) Priming and tolerance are also affected by the duration of the gap between two doses. Very long gaps fail to exhibit either priming or tolerance.

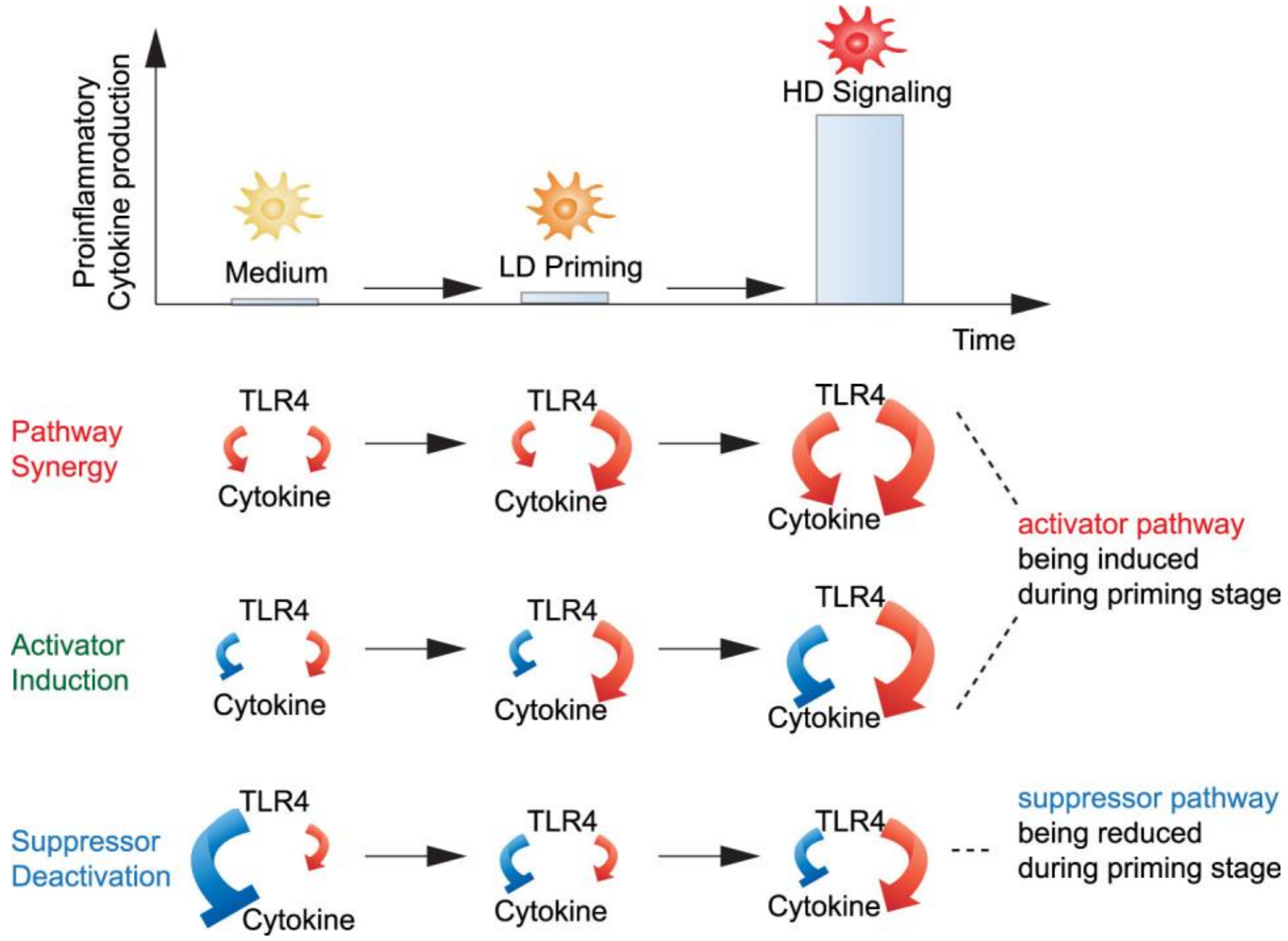

**Figure 7.** Schematic illustration of constructive (PS) and destructive (AI, SD) pathway interference leading to priming effect. PS results from the activation of the LD-responsive pathway ($x_2$) which cooperates with the other HD-responsive pathway ($x_1$) to boost cytokine expression in response to the following HD stimulus. AI results from activating a LD-responsive pathway ($x_2$), which cancels the inhibition coming from the other HD-responsive inhibitor ($x_1$) during the HD stage. SD results from deactivating a constitutively expressed suppressor ($x_1$) during the priming stage. Red line with arrow head: activation pathway. Blue line with bar head: inhibition pathway. Line width denotes strength of the pathway controlling the downstream cytokine expression.

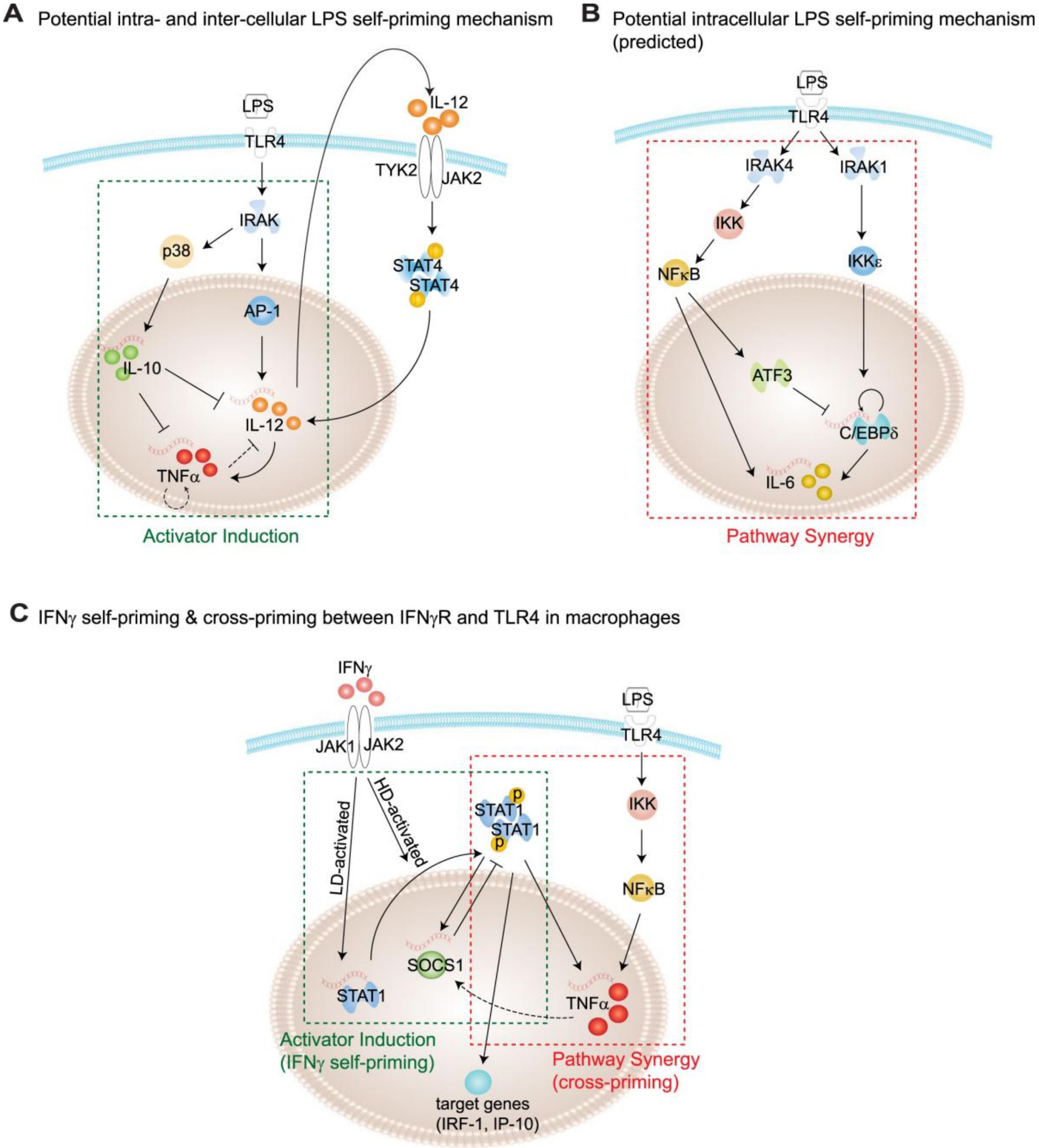

**Figure 8.** Example regulatory networks supporting the priming mechanisms. (A) The AI mechanism is consistent with observed intra- and inter-cellular molecular mechanisms for LPS priming, based on counterbalanced IL-10 and IL-12 signaling [19]. (B) The PS mechanism inspires this predicted intracellular molecular mechanism based on the selective activation of C/EBPδ by LD LPS. (C) IFN-γ self-priming and cross-priming to LPS follows the AI and PS mechanisms. Network details are retrieved from the database IPA (@Ingenuity) as well as the experimental literature listed in Table S3. Dashed lines refer to indirect regulations involving autocrine signaling loops.

# Supporting Information

## Network Topologies and Dynamics Leading to Endotoxin Tolerance and Priming in Innate Immune Cells

### Table of Contents

**Supporting text (Text S1):**

**Supporting figures:**

**Supporting tables:**

**Detailed criteria for priming and tolerance in the Metropolis searching algorithm**

We used the Metropolis algorithm [1] to search for parameter values for which the system exhibits priming or tolerance effects. Table S1 gives the criteria for identifying priming or tolerance parameter sets. In general, both priming and tolerance require the system to generate a dose-response curve having the following qualitative features: small signal (LD) gives small response and large signal (HD) gives large response; priming requires that LD+HD LPS gives a larger response than does a single HD LPS (positive control); tolerance requires that HD+HD LPS gives lower response than does a single HD LPS (positive control). Parameter sets that satisfy these conditions (either for priming or for tolerance) are called "good" sets.

**Two-stage Metropolis search for parameter sets that exhibit priming or tolerance**

It is impractical to perform a brute force search for priming/tolerance samples in a high dimensional parameter space. Figure S1A illustrates an alternative two-stage strategy. In the first stage, we searched widely over the parameter space with some bias to stay in a good parameter region and some chance to wander off in search of another good region. Then K-means Clustering and Principal Component Analysis was applied to the samples of good parameter sets generated in stage 1 to see if the data form several separate clusters. Each potential cluster provides a random seed for a second round of Metropolis searching. This time the search is restricted to stay within a good region, in order to search each region thoroughly and to obtain a representative sample of good parameter sets.

To apply the Metropolis Algorithm, we relate the current problem of searching in the parameter space to sampling the partition function of a pseudo-statistical physics system. The bias controlling the probability of wandering out of a good region ($\Omega_k = 0,\ \Omega_{k+1} = 1$) is defined by a Boltzmann-type expression $\rho = e^{-\beta(\Omega_{k+1} - \Omega_k)}$ where $\beta$ represents an "inverse temperature" variable. There exists a trade-off value of $\beta$ for the Metropolis search in stage I. If $\beta$ is too large, the search will stay in a local minimum and fail to explore the parameter space thoroughly. If $\beta$ is too small, the search cannot yield enough samples for the clustering analysis. Through trial and error, we found that $\beta$ = 6 is a good value for the stage I Metropolis search, which gives $\rho$ = 0.0025. Note that the priming region is very small compared to the whole parameter space. Therefore, although $\rho$ = 0.0025 is very small, it still guarantees that the system has sufficient probability to leave the good regions and thoroughly search the parameter space.

In the above procedure, the score function $\Omega_k$ plays the role of "energy" in a physical system. In general it can be a continuous function, and its gradient can guide the Metropolis search to the favorable region. For the current problem, the score function we use essentially behaves as a two-state system. Therefore we assign the value of $\Omega_k$ to be 0 or 1.

We chose to use the Metropolis method for the first stage, but other methods will probably work equally well, e.g. genetic algorithm [2] and the methods used by Ma et al. [3] and Yao et al. [4].

Figure S1B provides the result of the two-stage Metropolis search. In the left panel the priming sets obtained from the first stage form three main clusters under the K-means Clustering. For visualization purpose the clusters in the high-dimensional parameter space are plotted using the first two components of Principal Component Analysis. Using the Khachiyan Algorithm [5], we calculated the minimal volume ellipsoid to embrace 99% of the parameter sets of each region. As shown in the right panel of Figure 1B which calculates the distance of a parameter set to the center of each bounding ellipsoid, it turns out that a single ellipsoid embraces clusters 1 and 2, thus forming one single region (we call it "Region I"). This result is independently confirmed with the following Metropolis simulation with $\rho = 0$: a trajectory starting from one cluster can generate parameter sets belonging to the other cluster. On the other hand, cluster 3 forms a separate region (Region II). Notice that a small portion of samples locate within both ellipsoids, indicating these two ellipsoids (regions) are barely connected. We found that Region II is actually (part of) the mirror image of Region I with the roles of $x_1$ and $x_2$ exchanged, reflecting the symmetry of the 3-node system. Therefore, the results discussed below and in the main text focus on the motifs found in Region I.

About $10^6$ output samples are generated out of $10^8$ Metropolis steps in stage 2. Of these $10^6$ samples, some appear to be biologically irrelevant and are removed from the sample set. For example, in some cases $x_3(t)$ increases to a much higher level after the HD stimulation is removed, this would be a pathological response of the system. Other samples show unrealistically large sensitivity to initial conditions, i.e., although LD induced only small changes in $x_1$, $x_2$ and $x_3$ (less than 10%), the system still exhibited priming effect. If priming were due to such small differences, then (in our opinion) the response would not be robust to the stochastic fluctuation expected in real systems [6-9].

While the results reported in the main text are from one trajectory result, the procedure was repeated several times with random initial start of the searching in stage 1. Results analyzed from different trajectories agree with each other, confirming the convergence of our two-stage Metropolis searching procedure.

**Statistical method used to identify backbone motifs**

A backbone motif is defined to be the simplest motif (the fewest number of non-zero $\omega_{ji}$'s) that is shared by most of the priming/tolerance network structures in a particular region. A backbone motif must be able to generate a priming/tolerance effect by itself. Identification of backbone motifs helps to define the core mechanism of priming or tolerance. Figure S3 shows the statistical method used to obtain the backbone motifs for the pathway synergy group.

Step 1: calculate the mean of each interaction coefficient $\omega_{ji}$ among all samples of the group, and map the mean values into a topological matrix $\tau_{ji}$ (see Material and methods in the main text for the method of parameter discretization).

Step 2: for each $\omega_{ji}$ calculate its coefficient of variation (CV = standard deviation divided by |mean|). The value of CV measures the dispersion of the data along each parameter dimension. A large value of CV suggests that a link is not essential and should not be part of the backbone motif. Only links with CV < *CutOff* should be part of a backbone motif. For CV > *CutOff*, $\tau_{ji} = 0$ in the backbone motif.

Step 2.1: determine the optimal value of *CutOff*. As *CutOff* decreases, the corresponding motif becomes simpler and therefore more samples contain this motif. However, the motif is a backbone motif only if it gives priming by itself. Therefore, there exists an optimal *CutOff* value so that the corresponding motif has the simplest topology that is still able to generate priming for some specific parameter sets. In this case the optimal *CutOff* = 0.54 (see the right figure in Step 2.1 of Supplement Figure 3).

Step 2.2: compare each dimension in the CV matrix to this optimal *CutOff* value, and obtain the corresponding backbone motif.

Figure S4 shows 2D histograms of parameter distributions under each priming mechanism (PS, SD and AI). These histograms clearly highlight the corresponding backbone motifs. For example, for the 2D histogram shown in Supplemental Figure 4A, the PS data form clusters where both $x_1$ and $x_2$ activate $x_3$ (2nd figure), and $x_3$ feeds back negatively on $x_2$ (4th figure). Also $x_2$ shows significant auto-activation but $x_1$ does not (data spread out horizontally in the 5th figure); this is in line with the backbone motif where $x_1$ auto-regulation is not essential for priming. Similarly, $x_1$ exerts strong inhibition on $x_2$, whereas the regulation from $x_2$ to $x_1$ can be either negative, zero or positive (the 3rd figure), in line with the backbone motif where this regulation is missing. In addition, the 1st figure indicates that $x_1$ should change on a much faster time-scale than $x_2$. This is a dynamical requirement of pathway synergy in addition to the topological features as illustrated by the backbone motif.

**Motif density is more robust than frequency to variation in the topological cut-off**

To map from the continuous space of interaction coefficients $\omega_{ji}$ to the discrete space of network topologies $\tau_{ji}$, one must choose a cut-off value $\tau_0$ for mapping $\omega_{ji}$'s to −1, 0 or +1. We have chosen this cut-off $\tau_0$ (somewhat arbitrarily) to be 0.1. The simplest way to order these topologies from "more robust" to "less robust" is in terms of the number of parameter sets that map into each topology, i.e., the frequency of each topology in the total data set. However, we find that topology-frequency is sensitive to the choice of the cut-off value for $\omega_{ji}$. A better measure is topology density (Figure S6), defined as follows. The total volume of the 9-dimensional space of interaction coefficients is $2^9$, because each $\omega_{ji}$ can continuously vary over [−1, 1]. For a motif with $m$ non-zero $\tau_{ji}$'s, the volume of its subspace is $(1-\tau_0)^m = (0.9)^m$. The density of the motif is defined as the number of samples corresponding to this motif divided by the volume of its subspace.

In Supplemental Figure 6 we compared the two ways of ordering the topologies using the SD data set as an example. The figure shows how the rank of robustness of each topology changes due to 10%, 30% and 50% positive or negative variations from the original cut-off $\tau_0 = 0.1$. A point on the figure with coordinate (*x*, *y*) means that the rank of a given topology is *x* with $\tau_0 = 0.1$, but *y* with the varied $\tau_0$. Scattering from the diagonal indicates changing of the ranking due to $\tau_0$ variation. The density-sorted rank (top panel) is less sensitive than the frequency-sorted one (lower panel) to the change of $\tau_0$.

**2D parameter correlations demonstrate how parameter compensation affects topological robustness**

We calculated the correlation matrix of each priming mechanism from the corresponding samples. As can be seen from Figure S8A, some parameters show strong anti-correlations. For a pair of anti-correlated parameters, increasing one can be compensated by decreasing the other (or negatively increasing the other if the regulation is inhibition), so the overall dynamics remains (approximately) the same. This is because in the modeling equations,

$$\frac{dx_j(t)}{dt} = \gamma_j\left(\frac{1}{1+e^{-\sigma_j W_j}} - x_j(t)\right)$$

$$W_j = \sum_{i=1}^{3} \omega_{ji} x_i(t) + \omega_{j0} + S_j$$

the activation of species $x_j$ is dependent on the overall net input $W_j$. As $W_j$ sums inputs from all regulating nodes, a change in one parameter (e.g. $\omega_{j1}$) can be compensated by a change in a second parameter (e.g. $\omega_{j2}$) if the sum stays the same. Such parameter compensation expands the region of parameter space where priming or tolerance is observed and therefore affects the robustness of the model.

For example, the left panel of Supplemental Figure 8B shows that the feedback from $x_3$ to $x_2$ strongly anti-correlated with $x_2$'s auto-regulation among SD datasets. With $\omega_{23} = 0$, the absolute value of $\omega_{22}$ needs to be also small (the Null region in the right panel of Figure 8B), otherwise priming is abolished. However, since $\omega_{23}$ and $\omega_{22}$ are anti-correlated, the effect of an increasing $\omega_{22}$ can be canceled off by increasing $\omega_{23}$, thus expand the priming region in the parameter space (the upper left and bottom right regions of the right panel of Figure S8B).

## Supporting References

**Table S1.** Criteria identifying priming and tolerance for a given parameter set *x*.

| A Good set of | Single LD | Single HD | LD+HD | HD+HD |
|---|---|---|---|---|
| **Priming** | $R_{LD}(x) < \delta_{LD}$ | $R_{HD}(x) \geq \delta_{HD}$ | $R_{LD+HD}(x)/R_{HD}(x) \geq \lambda$ | - |
| **Tolerance** | $R_{LD}(x) < \delta_{LD}$ | $R_{HD}(x) \geq \delta_{HD}$ | - | $R_{HD}(x)/R_{HD+HD}(x) \geq \lambda$ |
| ***Description*** | LD signal stimulates small response. | HD signal stimulates large response. | Two sequential signals (LD followed by HD) gives a larger response than a single HD. | Two sequential signals (HD followed by HD) gives a smaller response than a single HD. |

*R* denotes the maximum response of "cytokine" $x_3$ under a specific stimulation protocol. LD: low dose; HD: high dose; LD+HD: LD followed by HD with maximum response measured in the HD period; HD+HD: HD followed by HD with maximum response measured in the second HD period. $\delta_{LD}$ and $\delta_{HD}$ denote the threshold of response under LD and HD, respectively. $\lambda > 1$ is the threshold of fold-change in the maximum response. The values we have chosen for these parameters (LD=0.1, HD=1, $\delta_{LD}=\delta_{HD}=0.3$, $\lambda=1.5$) are in qualitative agreement with experimental observations.

**Table S2.** Parameter sets used to generate time course and phase-space trajectory in Figure 3 and Figure S5.

| | **PS** | **PS** | **AI** | **AI** | **PS** | **PS** |
|---|---|---|---|---|---|---|
| | bistable | monostable | bistable | monostable | bistable | monostable |
| $\omega_{11}$ | 0.26 | 0.19 | -0.54 | 0 | 0.86 | 0.84 |
| $\omega_{12}$ | -0.92 | -0.27 | 0.05 | -0.11 | -0.78 | -0.90 |
| $\omega_{13}$ | 0.61 | 0.23 | -0.24 | 0.04 | -0.86 | -0.36 |
| $\omega_{21}$ | -0.95 | -0.93 | -0.61 | -0.52 | 0.36 | 0.08 |
| $\omega_{22}$ | 0.53 | 0.54 | 0.99 | 0.95 | 0.06 | 0.16 |
| $\omega_{23}$ | -0.54 | -0.35 | -0.69 | -0.89 | -0.53 | -0.45 |
| $\omega_{31}$ | 0.18 | 0.18 | -0.80 | -0.75 | -0.96 | -0.85 |
| $\omega_{32}$ | 0.47 | 0.27 | 0.83 | 0.82 | 0.89 | 0.93 |
| $\omega_{33}$ | 0.12 | 0.40 | 0.69 | 0.77 | 0.61 | 0.54 |
| $\gamma_1$ | 1.56 | 0.43 | 0.10 | 0.11 | 0.14 | 0.15 |
| $\gamma_2$ | 0.11 | 0.11 | 0.19 | 0.16 | 0.76 | 9.96 |
| $\gamma_3$ | 1.00 | 1.00 | 1.00 | 1.00 | 1.00 | 1.00 |
| $\sigma_1$ | 6.84 | 8.00 | 4.36 | 4.37 | 7.96 | 5.36 |
| $\sigma_2$ | 7.19 | 8.00 | 6.55 | 6.89 | 6.33 | 5.50 |
| $\sigma_3$ | 6.00 | 8.00 | 6.00 | 6.00 | 6.00 | 6.00 |
| $\omega_{10}$ | -0.75 | -0.50 | -0.07 | -0.22 | -0.10 | -0.15 |
| $\omega_{20}$ | -0.25 | -0.25 | -0.25 | -0.25 | -0.25 | -0.25 |
| $\omega_{30}$ | -0.50 | -0.50 | -0.50 | -0.50 | -0.50 | -0.50 |

**Table S3.** Experimental literature supporting the network details in Figure 8.

| **Figure 8 Panel** | **Source** | **Target** | **Regulatory Type** | **Reference** | **Comment** |
|---|---|---|---|---|---|
| A | TLR4 | IRAK | Activation | [1,2] | |
| A | IRAK | P38 | Activation | [3,4] | |
| A | P38 | IL-10 | Transcription | [5] | |
| A | IL-10 | IL-12 | inhibition | [6,7] | |
| A | IL-10 | TNFα | inhibition | [8,9] | |
| A | IRAK | AP-1 | Activation | [10] | |
| A | AP-1 | IL-12 | Transcription | [11,12] | |
| A | IL-12 | TNFα | Transcription | [9,13] | |
| A | TNFα | TNFα | Positive auto-regulation involving an autocrine loop | [14] | |
| A | IL-12 | IL-12 | Positive auto-regulation involving an autocrine loop | [15] | IL-12 auto-regulates itself through Jak/Stat pathway with STAT4 being the major transcription factor. |
| A | TNFα | IL-12 | inhibition | [16,17] | TNFα inhibits IL-12p40 through TNFα signaling pathway. |
| B | IRAK4 | IKK | Activation | [2] | |
| B | IKK | NFκB | Activation | [2] | |
| B | NFκB | ATF3 | Transcription | [18] | |
| B | ATF3 | C/EBPδ | Inhibition | [19,20] | |
| B | NFκB | IL-6 | Transcription | [20] | |
| B | IRAK1 | IKKε | Activation | [21] | |
| B | IKKε | C/EBPδ | Activation | [21] | Low dose LPS induces the expression of C/EBPδ |
| | | | | | through IRAK1 and IKKε. |
| B | C/EBPδ | C/EBPδ | Transcription | [20] | C/EBPδ can bind onto its own promoter to enhance the transcription. |
| B | C/EBPδ | IL-6 | Transcription | [20] | |
| C | IFNγ | STAT1 | Transcription | [22] | Low dose IFNγ elevates STAT1 transcription, but not STAT1 phosphorylation. |
| C | IFNγ | P-STAT1 | Activation | [22] | Phosphorylation of STAT1 is activated only under high dose IFNγ. |
| C | P-STAT1 | SOCS1 | Transcription | [22] | |
| C | SOCS1 | P-STAT1 | Inhibit | [22] | SOCS1 inhibits the phosphorylation and activation of STAT1. |
| C | P-STAT1 | IRF-1,<br>IP-10 | Transcription | [22] | |
| C | P-STAT1 | TNFα | Transcription | [23] | P-STAT1 may synergistically cooperate with NFκB to activate the transcription of TNFα. |
| C | TNFα | SOCS1 | Transcription | [24] | TNFα might be able to negatively feedback on P-STAT1 through enhancing the production of SOCS1. |

**Table S3 References**

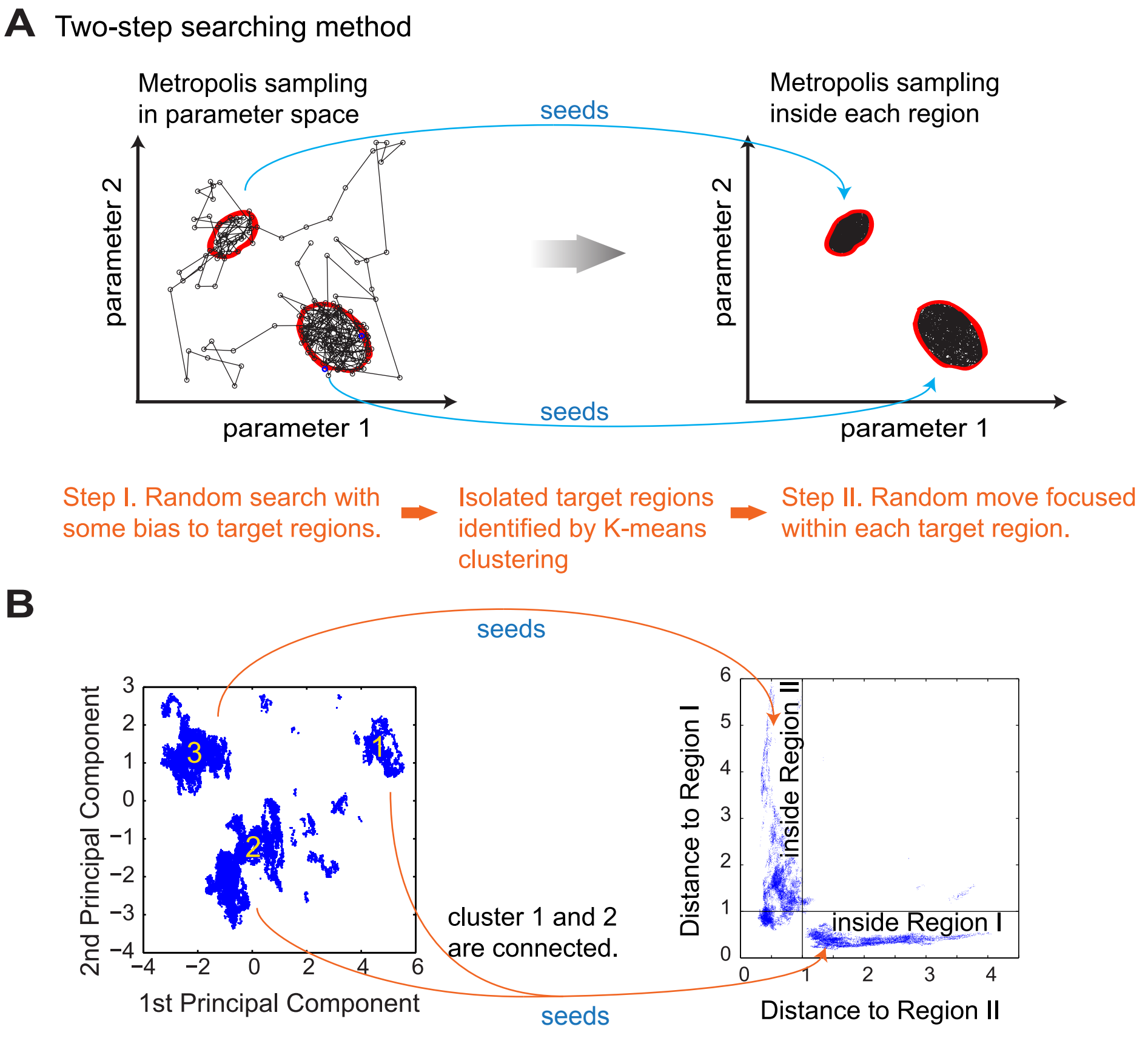

**Figure S1.** Illustration of the two-stage Metropolis search procedure. (A) Schematic illustration of the two-stage Metropolis search method for priming/tolerance parameter sets. In the first stage one randomly searches the whole parameter space. K-means clustering algorithm identifies one or more clusters of the data. Then one performs a second Metropolis step to search thoroughly inside each cluster. (B) As a result, we got three priming set clusters with K-means clustering. By calculating the minimum volume bounding ellipsoid, we found that cluster 1 and 2 belong to a single region (Region I) whereas cluster 3 belong to a separate region (Region II).

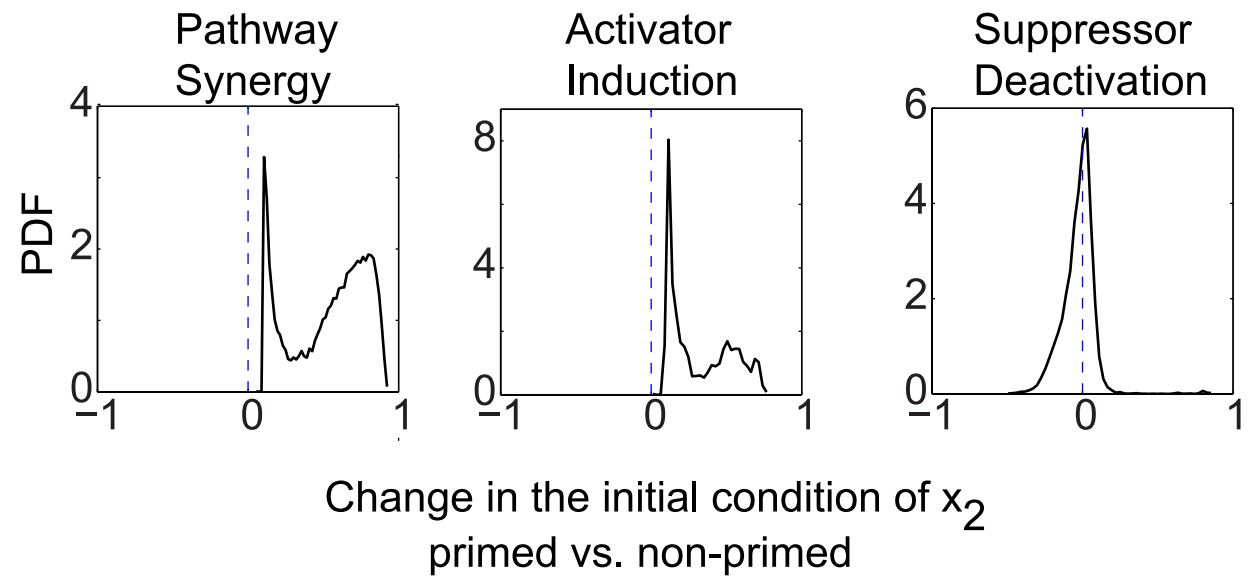

**Figure S2.** Distribution of change in x2's initial condition prior to HD without or without priming treatment. Both PS and AI show considerable increase in x2 in the primed system. PDF: probability distribution function.

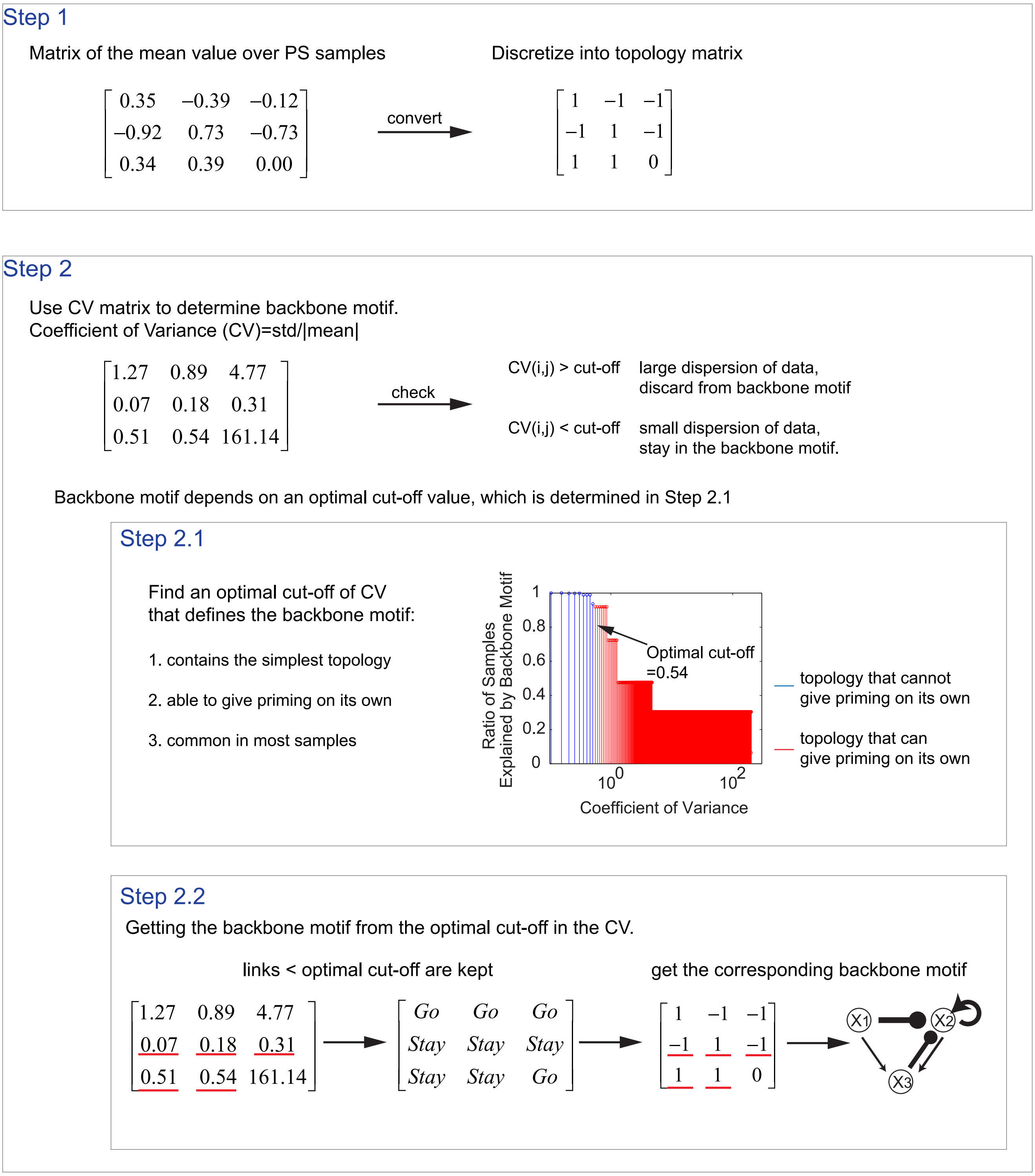

**Figure S3**. Statistical method used to identify backbone motifs from priming/tolerance data.

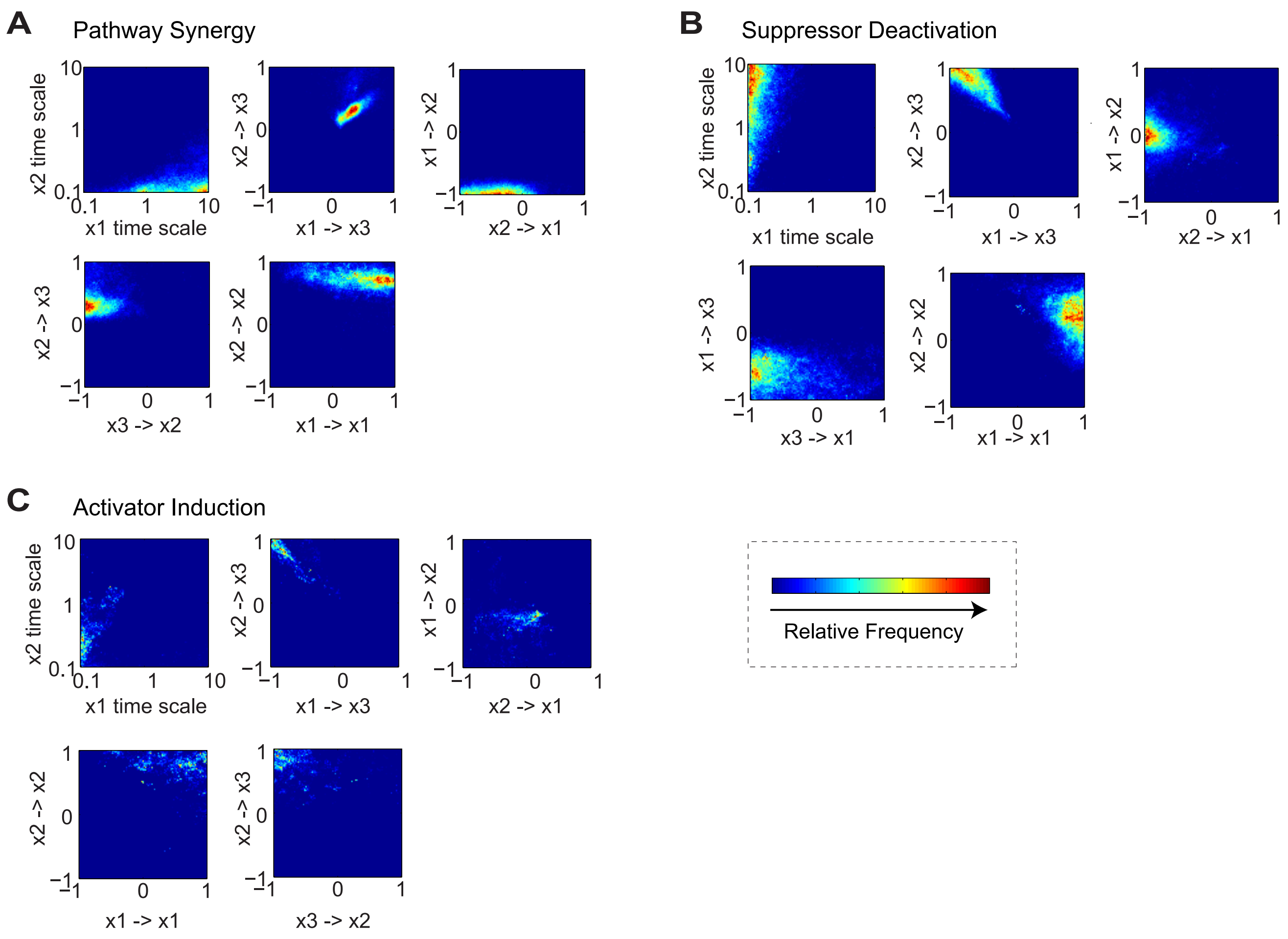

**Figure S4.** Parameter correlations highlight the backbone motifs of each priming mechanism. (A) Pathway Synergy, (B) Suppressor Deactivation, and (C) Activator Induction.

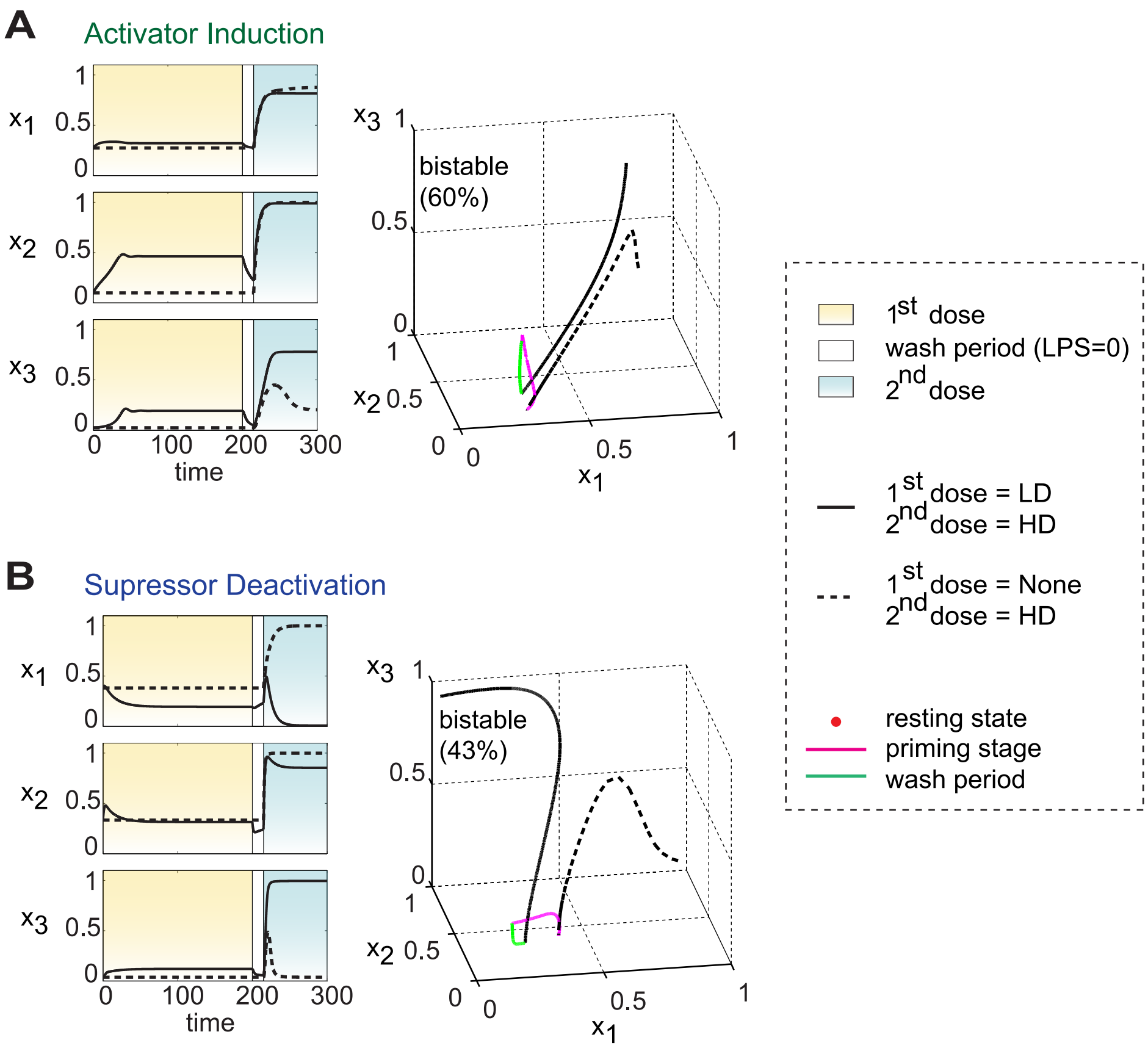

**Figure S5**. Typical time course and corresponding trajectory in the phase space. (A) bistable case of AI mechanism. (B) bistable case of SD mechanism. Refer to Figure 3 of the main text for the time course trajectories in other cases.

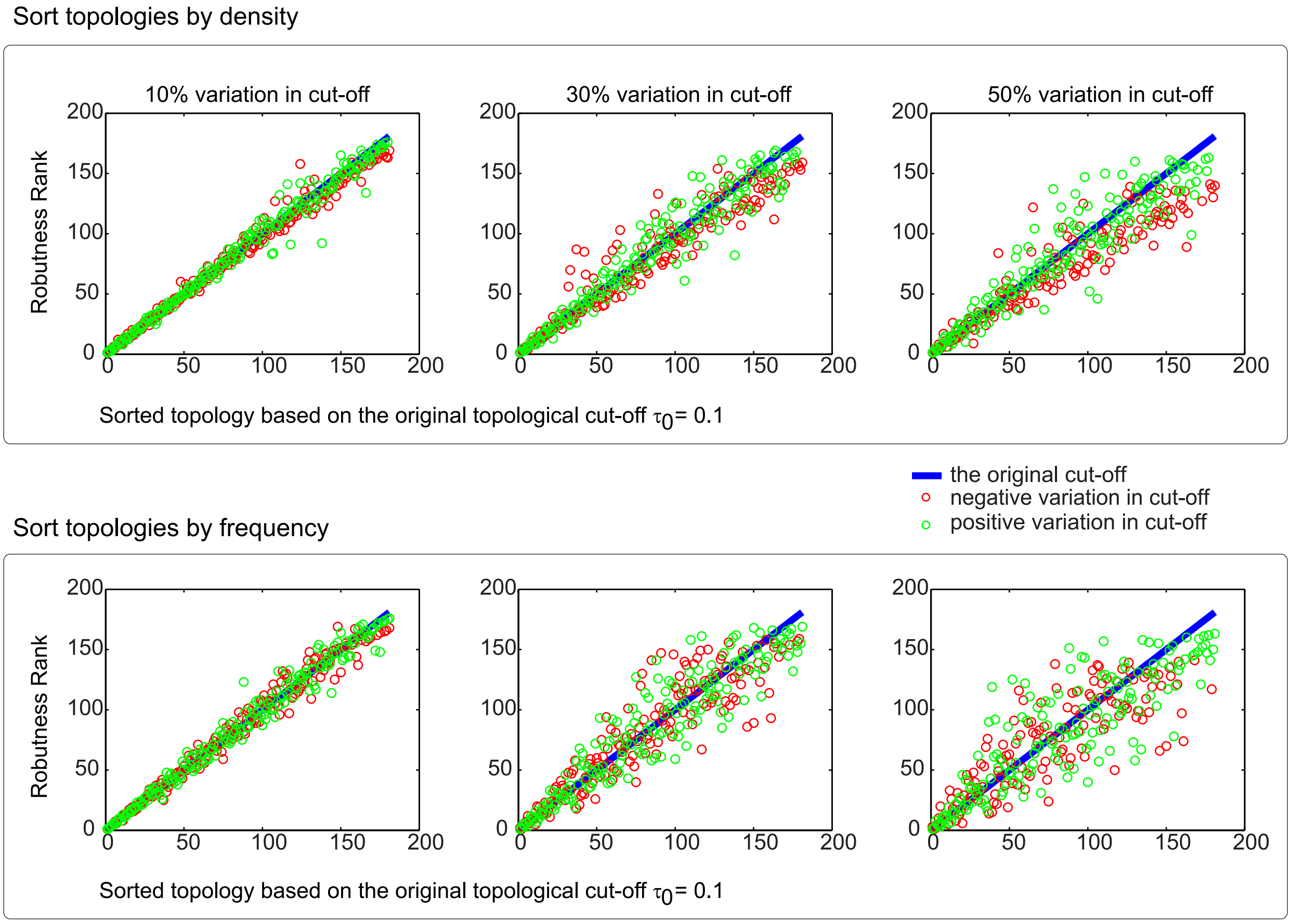

**Figure S6**. Change in the robustness rank as a result of variations in the topology cut-off. SD datasets are used as an example. The robustness rank is calculated based on density (top panel) or sample frequency (lower panel) of the unique topologies. Changes in the robustness rank is compared with 10% (left column), 30% (center column), and 50% (right column) variation in the topology cut-off $\tau 0=0.1$.

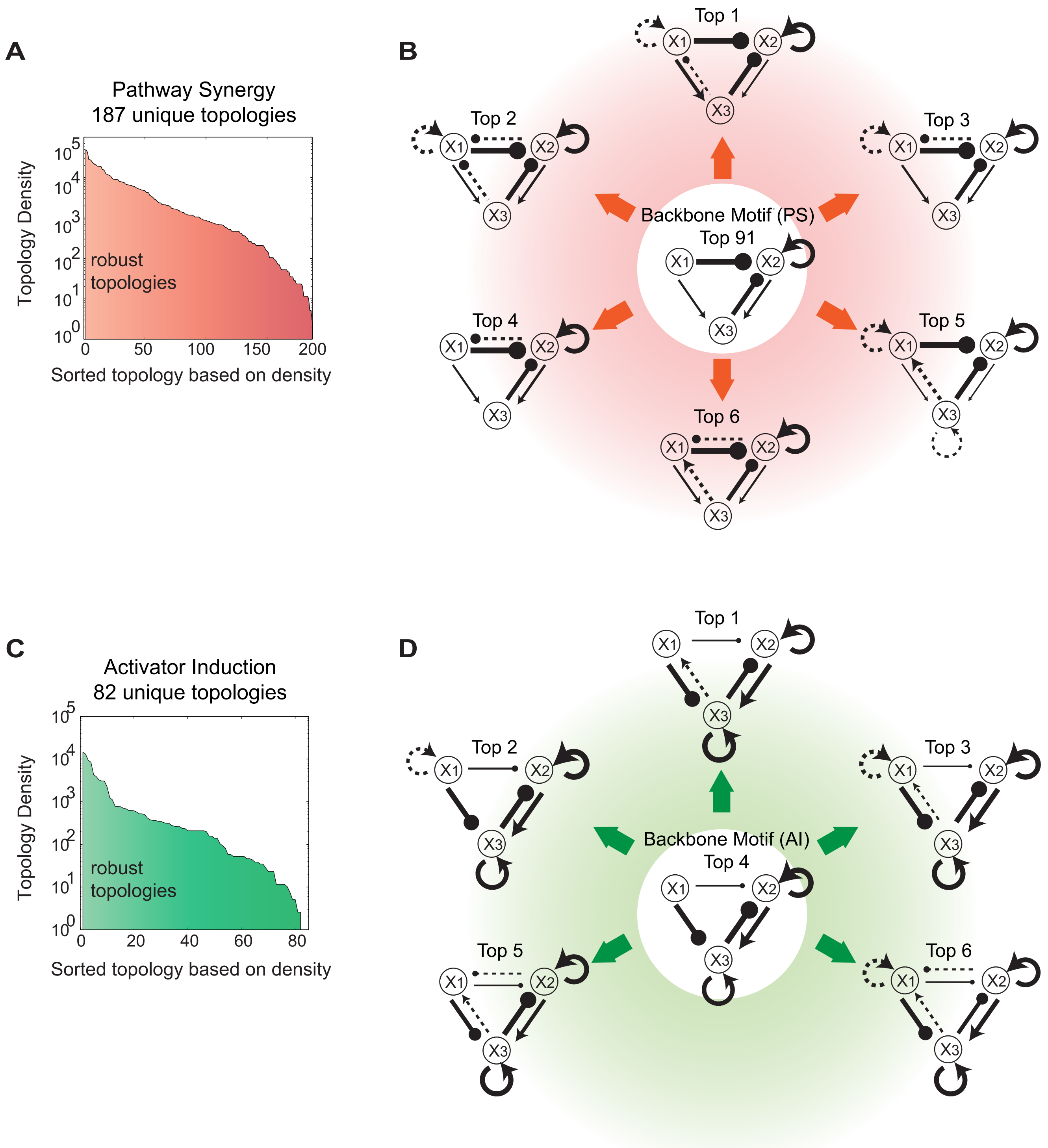

**Figure S7**. Topologies of the PS and AI mechanisms. (A) The topology density distribution for the PS mechanism. (B) Top six PS topologies and the backbone motif. (C) The topology density distribution for the AI mechanism. (D) Top six AI topologies and the backbone motif. Line widths are proportional to the mean value of samples of the corresponding topology. Dashed lines denote the additional links present in the top topologies but absent in the backbone motif.

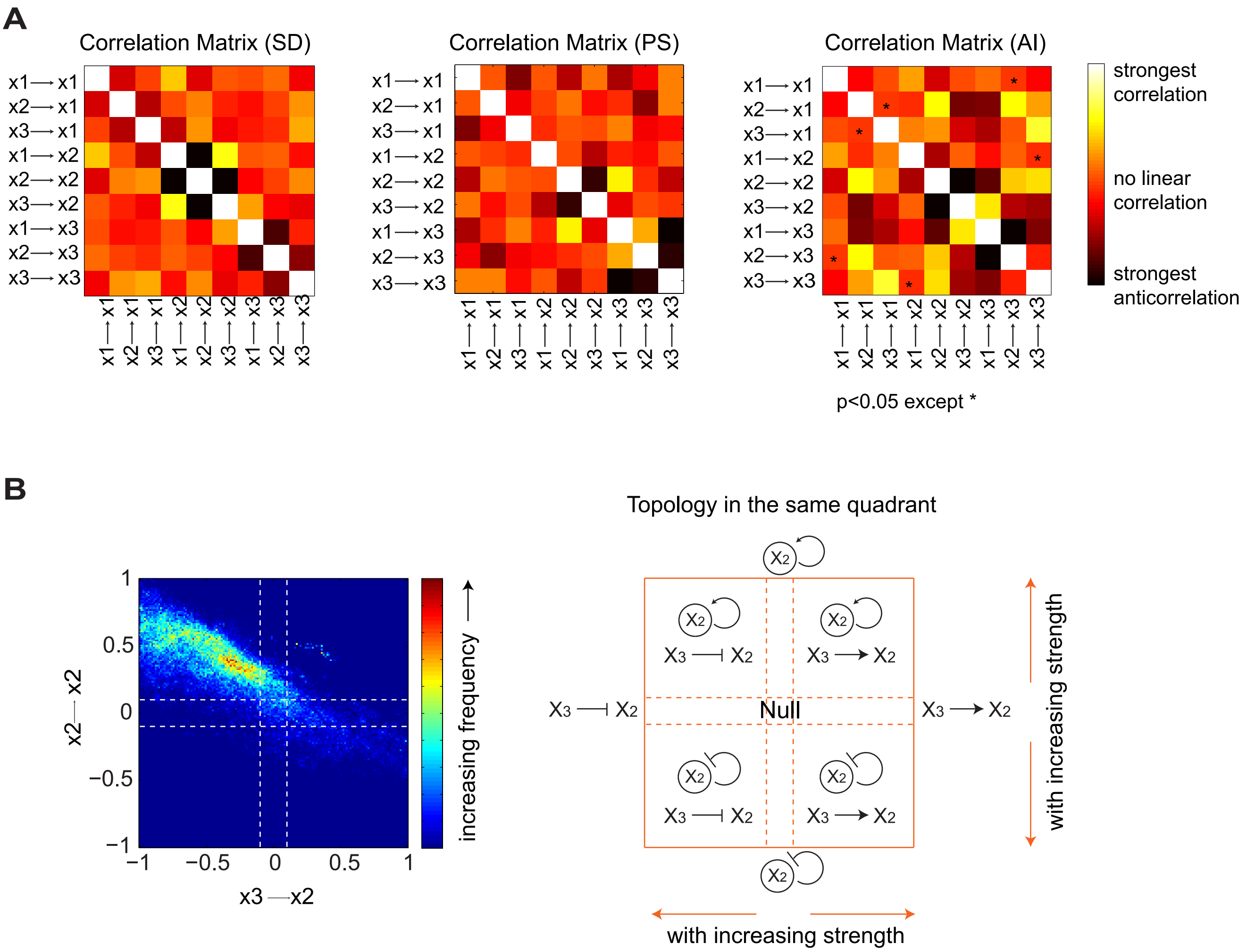

**Figure S8.** Parameter correlation and compensation affects the robustness of the model. (A) Correlation matrix calculated based on the samples of each priming mechanism. The p-value is smaller than 0.05 except where marked. (B) The parameter compensation mechanism is illustrated by the 2D correlation histogram of the SD samples (left) and the corresponding connection diagrams (right).